\documentclass[11pt]{aastex62}

\usepackage{amsmath}
\usepackage{amssymb}

\usepackage{graphicx}
\usepackage{epstopdf}
\usepackage{lineno}
\usepackage{ulem}
\newcommand{\eqs}[1]{\begin{equation}\begin{split}#1\end{split}\end{equation}}
\newcommand{\fermi}{{\it Fermi} }
\submitjournal{ApJ}
\accepted{\today}
\shorttitle{Population syntheses of millisecond pulsars
from the Galactic Disk and Bulge}
\shortauthors{Gonthier et al.}

\begin{document}


\title{Population syntheses of millisecond pulsars
from the Galactic Disk and Bulge}


\author{Peter L. Gonthier}
\affiliation{Hope College, Department of Physics, 27 Graves Place,  Holland, MI 49423, USA,
{\rm gonthier@hope.edu} }
,
\author{Alice K. Harding}
\affiliation{Astrophysics Science Division, NASA/Goddard Space Flight Center, Greenbelt, MD 20771, USA,
{\rm Alice.K.Harding@nasa.gov} }

\author{Elizabeth C. Ferrara}
\affiliation{Astrophysics Science Division, NASA/Goddard Space Flight Center, Greenbelt, MD 20771, USA,
{\rm Elizabeth.C.Ferrara@nasa.gov} }

\author{Sara E. Frederick}
\affiliation{University of Rochester, Department of Physics \& Astronomy, 211 Bausch \& Lomb Hall, Rochester, NY 14627-0171, USA}
\affiliation{University of Maryland, Department of Astronomy, 1113 Physical Sciences Complex, College Park, MD 20742-2421, USA,
{\rm sfrederick@astro.umd.edu} }

\author{Victoria E. Mohr}
\affiliation{Washington and Jefferson College, Department of Physics, 60 S Lincoln St, Washington, PA 15301, USA,
{\rm mohrve@gmail.com} }

 \author{Yew-Meng Koh}
 \affiliation{Hope College, Department of Mathematics, 27 Graves Place,  Holland, MI 49423, USA,
{\rm koh@hope.edu} }



\begin{abstract}
We present the results of a population synthesis of radio and $\gamma$-ray millisecond pulsars (MSPs) from the Galactic Disk (GD).  Using 92 radio millisecond pulsars detected in 13 radio surveys and 54 \fermi MSPs detected as point sources in the first point source catalog, we establish six free parameters corresponding to the overall factor and the exponents of the period and period derivative dependence for each of the radio and $\gamma$-ray empirical luminosity models.  We test three high-energy emission models described by the two pole caustic Slot Gap, Outer Gap, and Pair Starved Polar Cap geometries.  The simulated distributions of pulsar properties adequately describe the distributions of detected MSPs from the GD.  We explore the $\gamma$-ray emission from groups of MSPs in globular clusters and in the Galactic Bulge.  The simulation predicts reasonable numbers of \fermi MSPs detected in the other point source catalogs and anticipates a bright future for \fermi observations of MSPs, expecting a total of $\approx 170$ MSP detections from the GD within ten years.  Our numbers of simulated MSPs in globular clusters are in agreement with those derived from \fermi detections.  The simulation predicts about 11,000 MSPs in the Galactic Bulge are required to explain the $\gamma$-ray Galactic Center Excess. 
\end{abstract}





 \accepted{June 28, 2018}

\section{Introduction}
The study of $\gamma$-ray pulsars has recently undergone a major transformation with the launch of the {\it Fermi Gamma-Ray Space Telescope} (\fermi) in 2008.  In addition to an increase in number of known $\gamma$-ray pulsars from seven to currently more than 200\footnote{https://confluence.slac.stanford.edu/display/GLAMCOG/Public+List+of+LAT-Detected+Gamma-Ray+Pulsars}, \fermi has identified several different populations: young radio-loud pulsars, young radio-quiet pulsars and millisecond pulsars (MSPs).  The number of $\gamma$-ray MSPs (currently 96) is now approaching the combined total of young $\gamma$-ray pulsars (117), which at first may be surprising since the detected normal (non-recycled) pulsars (NPs) outnumber the MSPs, and MSPs are more efficient $\gamma$-ray emitters, converting greater amounts of rotational energy into $\gamma$ rays than NPs.   However, MSPs have on average higher spin-down power $\dot E_d$ than NPs and $\gamma$-ray luminosity is strongly correlated with $\dot E_d$.  \fermi has discovered $\gamma$-ray MSPs through several methods: using known radio ephemerides, through radio searches in unidentified point sources, and through $\gamma$-ray blind searches.  The radio searches of \fermi unidentified sources has been phenomenally successful (e.g. \cite{Ransom11}), with 85 MSPs having been discovered to date.  Most of these are found in areas of the sky not yet covered by sensitive radio surveys, but many have radio fluxes above those of the surveys \citep{Story07} (hereafter SGH).  The MSPs have a much broader distribution in Galactic latitude than NPs.   The diffuse $\gamma$-ray background at high latitudes is much lower than that concentrated toward the Galactic plane, making it easier to detect $\gamma$-ray MSPs as point sources and as $\gamma$-ray pulsars.  Almost all are radio-loud, whereas only about half of young $\gamma$-ray pulsars are radio-loud.  
The population of MSPs discovered in radio surveys has also been growing steadily, with advances in receiver technology and the effort to expand the number of MSPs for the Pulsar Timing Arrays that hope to detect low-frequency gravitational wave signals.  

These new observations of MSPs, particularly the $\gamma$-ray pulsations, have led to revised models of their high-energy emission.  Prior to \fermi launch, emission models assumed that because of their very low surface magnetic fields, most MSPs did not produce enough electron-positron pairs to screen the accelerating electric fields and restrict acceleration to small regions or narrow gaps in the pulsar magnetosphere.  Models thus focused on locations either near or above the magnetic polar caps \citep{Buli00, Lou00} or over the extended pair-starved open-field region (\cite{Hard05,Vent05} pair-starved polar cap models (PSPC)).  \cite{Zhan03} pointed out that formation of outer gaps in MSPs would require multipole fields near the surface.  The $\gamma$-ray spectra and light curves were therefore not expected to look the same as those of young pulsars.  However, \fermi data revealed that $\gamma$-ray MSP spectra and light curves are very similar to those of young pulsars, suggesting that their high-energy emission comes from narrow gaps in the outer magnetosphere.  Most MSPs must produce unexpectedly high pair multiplicity, possibly requiring non-dipolar field structure \citep{Zhan03, Hard11}.  Although a small number of MSP light curves are well described by PSPC models \citep{Venter09}, to account for the observed phase lags between their radio and $\gamma$-ray peaks their emission must come from altitudes well above the polar caps.  

Making use of both the recently expanded observations of MSPs and revised theoretical understanding of their emission, we have performed a new population synthesis.  Although our previous study (SGH) predicted fairly accurately the number of MSPs discovered in {\it Fermi} sources, this study assumed a PSPC model for the $\gamma$-ray emission flux and geometry.   Allowing for a wider range of emission geometries, including narrow outer magnetosphere gaps, using the observed number and spectra of $\gamma$-ray MSPs detected by \fermi and including new radio surveys, we can obtain an improved picture of the MSP population in the Galaxy.  Modeling the underlying population of Galactic MSPs is important for better understanding neutron star magnetic field evolution,  for studying the contribution of MSPs to the observed $\gamma$-ray excess at the Galactic center \citep{Ajello16} and the cosmic-ray positron excess \citep{Venter15}, and for predicting future detections by \fermi and {\it NICER}.

This study presents a new and updated population synthesis of radio and $\gamma$-ray MSPs using a select group of radio surveys and the 11 month \fermi first point source catalog \citep{Abdo1FGL} (1FGL) to tune six free model parameters associated with the empirical radio and $\gamma$-ray luminosities.  The synthesis does not consider the stellar evolution of the binary system, but rather treats only the evolution of the system from the end of the accretion spin-up phase, from the start of the rotation-powered pulsar spin-down evolution to the present time.  One of the main goals of this study is to test three high-energy emission models to see if signatures might exist that  allow one to differentiate among these models.  We then simulate MSPs over longer time periods to compare our predictions with detected MSPs in more recent point source catalogs and into the future of {\it Fermi}'s mission.  In addition, we assess the contribution of the population of MSPs from the Galactic Disk to the diffuse $\gamma$-ray background.   Finally, we test our $\gamma$-ray emission model by simulating MSPs within a group of globular clusters to compare our predictions with those estimated by \cite{Abdo10}.  Unique to this synthesis is the simulation of radio and $\gamma$-ray MSPs within the same original population with independent ``detection" prescriptions implemented within the code.  We assess the short-comings of the study and future plans for improvements.  We invoke a new  population of MSPs in the Galactic Bulge with the same characteristics as those in the disk to explore the contribution of such a population to the Galactic Center Excess (GCE) \citep{Ajello16}.

\section{Select group of detected radio and $\gamma$-ray millisecond pulsars}\label{sec:selgrp}
We assume a MSP birthrate of 4.5 per Myr (SGH), and simulate a group of evolved MSPs ten times larger than the number required by the birthrate to obtain smoother simulated histograms.
We select a group of thirteen radio surveys that have discovered 92 MSPs in the Galactic field.  The surveys cover the sky with different sensitivities as they involved a variety of telescopes and sampling intervals.  Having selected these groups of radio and \fermi MSPs, we can tune the free model parameters of the luminosity models by maximizing the likelihood obtained from the comparison of the observed and simulated 1-D histograms of directly detected MSP characteristics within a Markov Chain Monte Carlo code.  To establish the free parameters of the $\gamma$-ray luminosity model, we use the number of  \fermi $\gamma$-ray MSPs detected as point sources in the 1FGL catalog.  Thus, for a given set of parameters, the simulation may not predict the exact number of detected 92 radio and 54 \fermi MSPs, but the simulated numbers will be close as a result of maximizing the likelihood in the MCMC simulations.

We focus on a group of 54 $\gamma$-ray MSP that \fermi detected as point sources in the 11 month 1FGL that were later discovered as pulsars through use of known radio ephemerides or through pointed radio observations, requiring that these \fermi MSPs have some radio flux.  We assume that all Fermi unidentified sources in the 1FGL that are MSPs have been observed and discovered in dedicated radio observations, so that the MSP population in 1FGL is complete.  The number of MSPs in later Fermi source catalogs may not yet have reached completeness.  We, therefore, consider these as radio-loud \fermi MSPs. Furthermore we can simulate \fermi MSPs that are above the $\gamma$-ray point source threshold, but are radio-weak with fluxes below those of current pointed radio observations.  In order to compare our simulated \fermi MSPs with those detected, we require a minimum radio flux $S_{1400} = 30\ \mu$Jy and a minimum dispersion measure $DM = 2.5\ {\rm pc/cm^3}$.   The majority of the 40 MSPs in the Second \fermi Pulsar Catalog \citep{Abdo2PC} (2PC) were either seen as point sources in 1FGL or were previously discovered to be radio pulsars.  We exclude from consideration the $\gamma$-ray pulsar J1823-3021A in a globular cluster.  Subsequent to publication of the 2PC, additional MSPs have been discovered as associated point sources in the 1FGL, including an MSP discovered in a \fermi blind search \citep{Pletsch2012}.  A group of 11 \fermi MSPs from 1FGL has previously been detected in radio surveys, corresponding to those listed in Table 3 of \citep{Abdo2PC} with the history type containing the letter ``r" excluding   J0030$+$0451,  J0218$+$4232,   J0751$+$1807, and J1614$-$2230 that were detected in miscellaneous surveys.  The pulsar J0030$+$0451 was discovered in the 430 MHz Arecibo drift scan survey \citep{Lommen00}, while J0218$+$4232 was identified as a steep spectrum, high polarized source with the Westerbork Synthesis Radio Telescope and discovered later as a pulsar with the Jodrell Bank telescope \citep{Navarro95}.  The remaining two pulsars, J0751$+$1807, and J1614$-$2230, were discovered in radio pointed observations of EGRET point sources \citep{Lundgren95, Crawford06}.  The assumption we are making here is that all MSP candidates in the 1FGL have been discovered as radio pulsars.  However, there may be as many as ten more undiscovered \fermi MSPs to be found in the 1FGL, introducing a systematic uncertainty in our predicted numbers.

\begin{figure}[ht!]
\begin{center}
\includegraphics[trim=0in 0in 0in 0in, width=5.in]{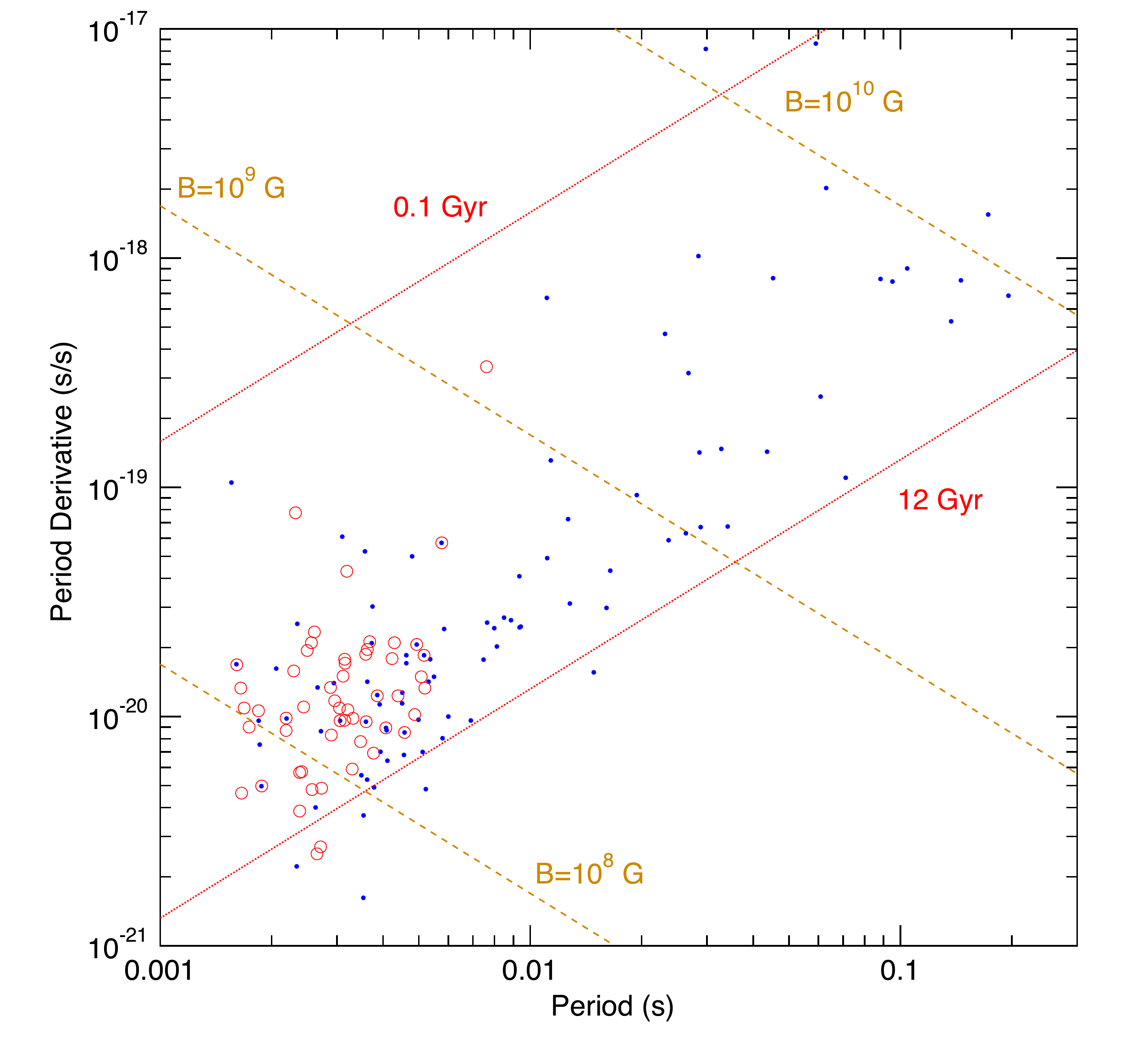}
\caption{$\dot{P}-P$ diagram of 92 radio millisecond pulsars (blue solid dots) in the ATNF catalogue
(http://www.atnf.csiro.au/people/pulsar/psrcat/ and \citet{Manchester05}) detected by the group of twelve radio surveys 
used in this study (see text) and 54 \fermi $\gamma$-ray MSPs detected as point sources in 1FLG (open red circles) detected
previously as radio pulsars in the select group of thirteen surveys.  Lines of constant characteristic age are indicated as red dotted lines, while the lines of 
constant magnetic field are indicated in brown dashed lines. }
\label{fig:RadFer}
\end{center}
\end{figure}

In addition to the ten radio surveys incorporated in the study of SGH, we have included the Parkes High-Latitude Multibeam Pulsar Survey \citep{Burgay06} and the Low- and Mid-Latitude Parkes High-Timing Resolution Surveys \citep{Keith10}.  This group of surveys contains 88 pulsars with periods less than 30 ms.  Removing the group of pulsars that do not have a measured $\dot P$, pulsars with negative $\dot P$ and pulsars in globular clusters leaves 76 short period pulsars.  However, we define the group of MSPs by the expression $\log(\dot P) < -19.5-2.5 \log (P)$ \citep{Acero15}, which results in a group of 92 detected radio MSPs.
These radio and $\gamma$-ray MSPs are presented in the $\dot P-P$ diagram shown in Figure (\ref{fig:RadFer}) as solid blue dots and open red circles, respectively.   The characteristic age of a pulsar $\tau_{\rm age} = P / (2\dot P)$ (red dotted lines) assumes the MSPs are born, after their accretion phase, with a zero period, which is clearly not necessarily the case as some MSPs have characteristic ages greater than the age of the universe.  In addition, the binary system requires a significant amount of time to evolve into a MSP through spin-up by mass accretion from a companion.  The lines of constant magnetic field indicated in Figure (\ref{fig:RadFer}) are calculated assuming that MSPs are orthogonal rotators, and all have a radius of $12$ km and  a mass of $1.6\, M_\odot$, where $M_\odot$ is a solar mass.  Perhaps it can be debated that the high-field MSPs are indeed normal pulsars as they have periods significantly greater than 30 ms.  However, the vast majority of this group are binary pulsars that clearly seem to belong to the valley of recycled MSPs.  As a result, our select group of \fermi pulsars consists of 54 radio-loud detected as point sources in 1FGL as well as the group of 92 radio pulsars detected in thirteen radio surveys.  

With the limited statistics of 54 \fermi and 92 radio detected MSPs, we have binned the data into 9 1-D histograms  (4 for radio and 5 for \fermi MSPs) of 10 bins each of the logarithms of the following pulsar characteristics: period, period derivative, radio flux ($S_{\rm 1400}$), dispersion measure, and $\gamma$-ray energy flux.
The observational uncertainties of these pulsar characteristics are much smaller than the width of the bin with the exception of the radio and $\gamma$-ray fluxes.  Since we are binning the logarithms of these MSP features, the width of the bin corresponds to a fixed fraction of the uncertainty of the characteristic divided by $\ln 10$.  The averages  of the fractional uncertainties (for example $\sigma_{S_{1400}}/S_{1400}$) of the radio fluxes $S_{\rm 1400}$ are 0.36, and for the \fermi $\gamma$-ray energy fluxes are 0.13 for the selected radio and \fermi MSPs.  The bin widths of the 1-D histograms of the radio and $\gamma$-ray fluxes are 0.4 and 0.185, therefore the full widths of the bins represent constant fractional uncertainties of 0.92 and 0.46, respectively.  Even half the bin size accommodates the fractional uncertainty of observed fluxes.  Therefore, we expect that the shapes of the 1-D histograms of the fluxes are well-defined and representative of the data.

\section{Present-day distribution of MSPs}  \label{sec:evolved}
The birth distributions of MSPs assumed in this study are those given by the work of \cite{Pacz90} with a radial scaling of 4.5 kpc and a scale height of 200 pc instead of 75 pc used in that work.  In addition, the supernova kick velocity model implemented was the same as in SGH with a Maxwellian distribution of width of 70 km/s (SGH).  These distributions generated the equilibrated distribution used in SGH.  The population synthesis does not consider the stellar evolution of the binary system, but rather a birth on the spin-up line at the time at which the accretion phase terminates, with an initial period given below.  We assume a constant birth rate of 4.5 per Myr and assign each of 450,000 MSPs a random age over the past 10 Gyr, evolving their trajectories within the Galactic potential \citep{Pacz90} from their birth location to their present location.  The code spatially evolves the group of MSPs, treating the binary system as a point particle within the Galactic potential that receives a kick velocity from the supernova forming the neutron star.  The velocity components are transformed to the Sun-centered coordinate system to obtain the proper motion of the MSPs, which was added to the intrinsic $\dot P$ of the simulated MSPs to compare with the observed $\dot P$ that do not have this Shklovskii effect correction \citep{Shklovskii1970}.  We simultaneously evolve the periods of the MSPs using the spin-down formulae given below.  We find that a variation of 20\% in the birth parameters does not significantly alter the final simulated histograms of MSPs characteristics, suggesting that the present-day spatial and velocity distribution of MSPs do represent an equilibrated present-day distribution as concluded by the work of SGH.  The group of 450,000 MSPs within the past 10 Gyrs corresponds to ten times the number required by the assumed birth rate (SGH).  We do so to improve the simulated statistics.  The group is evolved once, and the information is stored in a computer file for use with subsequent computer codes.

\subsection{Birth magnetic field and period distributions}
We assume that the magnetic field of a MSP does not decay with time.  As in the case of the study of SGH, the simulated histograms of pulsar characteristics agree well with those detected using a power law distribution of $B_8$ (in units of $10^8$ G) with a spectral index of $\alpha_{\rm B}$, having a normalized distribution given by the expression
\eqs{ \label{eq:Bf}
P\left(B_8\right) \; = \; & \frac{(\alpha_B+1)\,B^{\alpha_B}_8}{ B_{\rm max}^{\alpha_B+1}-B_{\rm min}^{\alpha_B+1}},
}
where $B_{\rm max} = 10^3$.  These parameters are fixed at the optimum values of $\alpha_B=-1.3$ and $B_{\rm min}=0.9$.  In the study of SGH a preferred index of $\alpha_B =-1$ and a $B_{\rm min}=1.0$ were used.

We assume a distribution of mass accretion birth lines from the Eddington critical mass accretion rate to about $10^{-3}$ of the critical value following the study by \cite{Lamb2005}.  We parameterize the mass accretion rates with a line in the $\dot P-P$ diagram as was done with Eq. (5) of SGH.  Eq. (6) of SGH is given by
\eqs{\label{eq:Po}
P_o \; = \; & 0.18\times10^{3\delta/7}\, B^{6/7}_8\ {\rm ms} ,}
and allows for the determination of the initial period with the restriction that $P_o\,>\, 1.3\ {\rm ms}$.  The study of SGH used a ramp distribution of $\delta$ characterized by a linearly increasing function with $\delta$.  We found improved agreement  by uniformly random dithering parameter $\delta$  between 0 and 2.    With these initial distributions and spinning down the MSPs, we can define a present-day spatial, magnetic field, and period distribution of MSPs that have then been used to find the contribution MSP to the cosmic ray background of positrons \citep{Venter15}.

\subsection{Pulsar spin-down}
Recently significant progress has been made in describing pulsar magnetospheres that are more realistic than the retarded vacuum dipole \citep{Deutsch55}.  Force-Free Electrodynamic solutions were obtained by \cite{Spitkovsky06} leading to the description of the pulsar spin-down by an expression similar to
\eqs{\label{eq:Spit}
L_{\rm sd}\; \sim \; & \frac{\mu^2\, \Omega^4}{c^3}\left(1+\sin^2\alpha\right), }
where $\mu$ is the magnetic dipole moment, $\Omega$ is the rotational angular velocity, $c$ is the speed of light, and $\alpha$ is the magnetic inclination angle relative to the pulsar's rotational axis.  \cite{Li12} constructed solutions to magnetospheres filled with resistive plasma, arriving at a very similar spin-down formula.   \cite{Conto14} considered the ideal force-free magnetosphere everywhere except within a current layer and arrived at a similar prescription for the pulsar spin-down luminosity.  Such results encouraged us to implement such a spin-down model into our population synthesis code.  Using a dipole moment of $\mu=B_o\,R^3\,/\, 2$, where $R$ is the stellar radius, and equating the spin-down luminosity to the rate of rotation energy loss $\dot E$, yields the expression
\eqs{\label{eq:Spin}
B^2_o\; =\; & \frac{c^3 I P \dot P}{\pi^2\,R^6\left(1+\sin^2\alpha\right)} } 
Integrating this equation over the age of the pulsar $ t $ provides the expression for obtaining the present-day period $P$ from the pulsar's birth period $P_o$
\eqs{\label{eq:PresP}
P^2\; =\; &P_o^2 +  \frac{ 2 \pi^2\,R^6}{c^3 I } \left(1+\sin^2\alpha\right)\,B_o^2\, t  } 
We assume $R\,=\,12\ {\rm km}$ and $M\,=\,1.6\ M_\odot$.  We use the prescription outlined in Section 2 of \cite{Pier12} to obtain the moment of inertia, which with these values of $R$ and $M$ yields a moment of inertia of $I = 1.7 \times 10^{45}\  {\rm g\cdot\,cm^2}$.  While there is growing evidence that the inclination angle becomes aligned with the neutron star's rotational axis with time in the case of normal pulsars, we do not consider such an alignment model in the case of MSPs in this study as there is currently no evidence for an alignment of the inclination angle with age for MSPs.  We assume that the supernova explosions do not select a particular distribution of inclination angle $\alpha$ therefore we randomized the $\alpha$ distribution uniformly between $0$ and $90^\circ$ whereas the distribution of viewing angle $\zeta$ was chosen to be uniformly distributed on the surface of a sphere.  We find that randomizing the inclination angle $\alpha$ in a similar manner as $\zeta$ leads to very similar results with no noticeable signature to distinguish between the models, therefore we prefer a uniform distribution.

\section{Radio emission} \label{sec:radio}
We implement a single cone and core beam geometry that has been extensively described in \cite{Hard07} simulating normal pulsars and in SGH simulating MSPs.
We use the spectral indices of -1.72 and -2.36 for the cone and core beams, respectively, that were also used in SGH.  We implement a slight modification to the ratio of core-to-cone peak flux model by converting the model into a ratio of core-to-cone luminosity model involving a broken power law with the approximate form
\eqs{\label{eq:LumRC}
{\cal R}\;=\; &\left\{ \begin{array}{l} 6.2\,P\,\dot P_{-15}^{-0.07}\, ,\ \ \ P < 0.7\,{\rm s}\,, \\
P^{-2.1}\,\dot P_{-15}^{-0.07}\, ,\ \ \ P \geq 0.7\,{\rm s}\, . \\
\end{array} \right. }
In terms of the total radio luminosity $L_\nu$, the core and cone luminosities are given by
\eqs{\label{eq:CrCn}
L_{\rm core}\; =\; & \frac{{\cal R}\, L_\nu}{1+{\cal R} } \\
 L_{\rm cone}\; =\; & \frac{ L_\nu}{1+{\cal R} }. }
Assuming that radio pulsars are standard candles, we incorporate a luminosity model similar to the one first used in the study of \cite{Arzo02}.  However, we investigate the exponents of $P$ ($\alpha_\nu$) and $\dot P$ ($\beta_\nu$), along with the overall factor $f_\nu$ by treating them as free parameters in a Markov Chain Monte Carlo (MCMC) model described later in Section \ref{sec:MCMC}.  The total luminosity is parameterized with the expression
\eqs{\label{eq:LumT}
L_\nu\;=\;  6.625\times 10^4\, f_\nu\, P^{\alpha_\nu}_{-3} \dot P^{\beta_\nu}_{-21} \ \ {\rm mJy\cdot kpc^2\cdot MHz}, } 
where the period derivative $\dot P_{-21}$ is in units of $10^{-21}\, {\rm s/s}$ and the period $P_{-3}$ is in units of $10^{-3}\, {\rm s}$.  With the luminosity model expressed in this manner, the $f_\nu$ parameters only needs to vary by a few.
This model was used in the previous studies of \cite{Gont04} and \cite{Hard07} where a population study of normal pulsars was performed with the exponents $\alpha_\nu=-1.3$ and $\beta_\nu=0.4$.  These two studies differed in the description of the radio beam geometry.  In the work of SGH, it was stated that a radio luminosity with $\alpha_\nu=-1.05$ and $\beta_\nu=0.37$ was able to adequately account for both NPs and MSPs.  To preserve the birthrate, we take the entire set of evolved MSPs with $\alpha_\nu$, $\beta_\nu$, and $f_\nu$ as free parameters. 

As discussed in Section 3 of \cite{Hard07}, we describe the radio flux of the simulated pulse profile for a survey at a frequency $\nu$ with the expression
\eqs{\label{eq:RaFx}
S\left(\theta,\nu\right)\;=\;& F_{\rm core}\,e^{-\theta^2/\rho_{\rm core}^2}+F_{\rm cone}\,e^{-\left(\theta-\bar\theta\right)^2/w_e^2}, }
where
\eqs{\label{eq:Flux}
F_i\left(\nu\right)\;=\;&\frac{-\left(1+\alpha_i\right)}{\nu}\left(\frac{\nu}{\rm 50\ MHz}\right)^{\alpha_i+1}\, \frac{L_i}{\Omega_i\, d^2} ,}
and $i$ refers to the core or cone with spectral index $\alpha_i$ and luminosity $L_i$, and $d$ is the distance to the pulsar in kpc.  The polar angle $\theta$ relative to the magnetic axis is given by the standard expression
\eqs{\label{eq:Theta}
\cos\theta\;=\;& \sin\alpha\sin\zeta\cos\phi+\cos\alpha\cos\zeta,}
where $\alpha$ in the inclination angle of the magnetic axis relative to the rotational axis, and $\zeta$ is the viewing angle relative to the rotational axis.  The factors $\Omega_i$ are not really the solid angles of the core and cone beams as the pulse profile is a pencil beam at a viewing angle $\zeta$ intersecting the emitting surface.  Rather, they are normalization factors given by 
\eqs{\label{eq:Norms}
\Omega_{\rm core}\; =\; & \pi\rho^2_{\rm core}, \\
\Omega_{\rm cone}\;=\;& 2\pi^{3/2}\,w_e\,\bar\theta, }
where the angular distribution of the intensity is normalized to unity.
The effective width of the core beam is given by the commonly used expression
\eqs{\label{eq:CrWd}
\rho_{\rm core}\;=\;& 1.5^{\rm o} P^{-0.5},}
similar to the one used by \cite{Rankin90}.
The cone beam is assumed to be a hollow Gaussian beam with an annulus and width given by the forms
\eqs{\label{eq:Cnbe}
\bar\theta\;=\;&\left(1-2.63\delta_w\right)\rho_{\rm cone}, \\
w_e\;=\;& \delta_w\rho_{\rm cone} , }
where $\delta_w=0.18$ \citep{Gont06}, and the apparent pulse width (see Eq. (3) of \cite{Kijak98})
\eqs{\label{eq:CnRho}
\rho_{\rm cone}\;=\;& 1.24^{\rm o}\, r_{\rm KG}^{0.5}\,P^{-0.5}.}
In order to take into account the observed radius-to-frequency mapping of the cone component in radio pulsars, we use the emission altitude $r_{\rm KG}$ given by the prescription given by Eq. (3) of the work of \cite{Kijak03} 
\eqs{\label{eq:rKG}
r_{\rm KG}\;=\;& 40\, {\dot P_{-15} } ^{0.07}\,P^{0.3}\,\nu_{\rm GHz}^{-0.26} ,}
where the altitude of emission is given in stellar radii and $\nu_{\rm GHz}$ is the survey frequency in gigaHertz.  The polar angle relative to the magnetic axis of the last open-field surface at a distance $r$ in stellar radii from the stellar center is
\eqs{\label{eq:ThOP}
\sin\theta_{\rm ofs}\; =\; & \left(\frac{r}{R}\right)^{0.5}\sin\theta_{\rm PC}, }
where $R$ is the stellar radius and the polar cap angle is 
\eqs{\label{eq:ThPC}
\sin^2\theta_{\rm PC}\;=\;&\frac{2\pi R}{P c} ,}
where $c$ is the speed of light.  Assuming small angles and $R=12\ {\rm km}$, the last open-field surface angle is 
\eqs{\label{eq:OFS}
\theta_{\rm ofs}\; =\; & 0.91^{\rm o}\, r^{0.5}\,P^{-0.5} ,}
compared to
\eqs{\label{eq:Thbar}
\bar \theta \;=\; & 0.65^{\rm o}\, r^{0.5}_{\rm KG}\, P^{-0.5}, \ {\rm and} \\
w_e\;= \; & 0.22^{\rm o}\, r^{0.5}_{\rm KG}\, P^{-0.5}, }
which suggests that the peak of the conal emission is about one width $w_e$ from the last open-field surface at $\theta_{\rm ofs}$ with about 15\% of the emission taking place within the closed-field region.

While the pulse profile of MSPs may be distorted due to aberration and the sweep back of the field lines from the retarded potential, the selection criterion, where the phase-averaged flux is above the minimum flux threshold $S_{\rm min}$ of a particular radio survey, is not very sensitive to distortions of the profile.  However, we  define a duty cycle ${\cal D}$ of the profile as the equivalent width of the pulse profile divided by the period, or by 
\eqs{\label{eq:Equiv}
{\cal D}\;=\;& \frac{ \int\limits_0^{2\pi} F\left(\zeta,\phi\right) d\phi} {2 \pi \, F_{\rm max}}, }
where $F\left(\zeta,\phi\right)$ is the flux profile of the pulsar at a given viewing angle $\zeta$ and as a function of the phase angle $\phi$, and $F_{\rm max}$ is the maximum intensity of the profile. 
If the profile of a pulsar is too broad ${\cal D}_{\rm max}>60\%$, it is unlikely to be detected as a radio pulsar.  Pulsars with narrow pulse profiles or small duty cycles are more readily detected.  

The radio emission and geometries of the core and cone beam were developed for NPs in the study of \cite{Gont04} and later the geometry evolved to the present one in the NP study of \cite{Hard07}.  The same model was applied to MSPs in the study of SGH.  The present model was used in the study of radio and $\gamma$-ray light curves for the case of NPs in the work of \cite{Pier12} and for the case of MSPs in the works of \cite{Venter09, Venter12} and of \cite{Johnson14} with the conclusion that the radio emission and beam geometry models worked well for both NPs and MSPs.

\section{Gamma-ray emission}
We chose to implement an empirical model of the $\gamma$-ray luminosity $L_\gamma$ (from both poles) in a similar fashion as the one implemented in the radio luminosity in Eq. (\ref{eq:LumT}) with the form
\eqs{\label{eq:GamLum}
L_\gamma\;=\; 1.724\times 10^{34}\, f_\gamma\,P^{\alpha_\gamma}_{-3}\,\dot P^{\beta_\gamma}_{-21}\ \ {\rm eV \cdot s^{-1}} .} 
The simulation is normalized to the assumed birthrate, while we consider $\alpha_\gamma$, $\beta_\gamma$ and $f_\gamma$ (varying within a small range) to be free parameters, searching within the parameter space of the model for the values of both $\gamma$-ray and radio luminosity models discussed in Section \ref{sec:radio} that maximize the likelihood.

\begin{figure}[ht!]
\begin{center}
\includegraphics[trim=1in 0in 0in 0in, width=6.in]{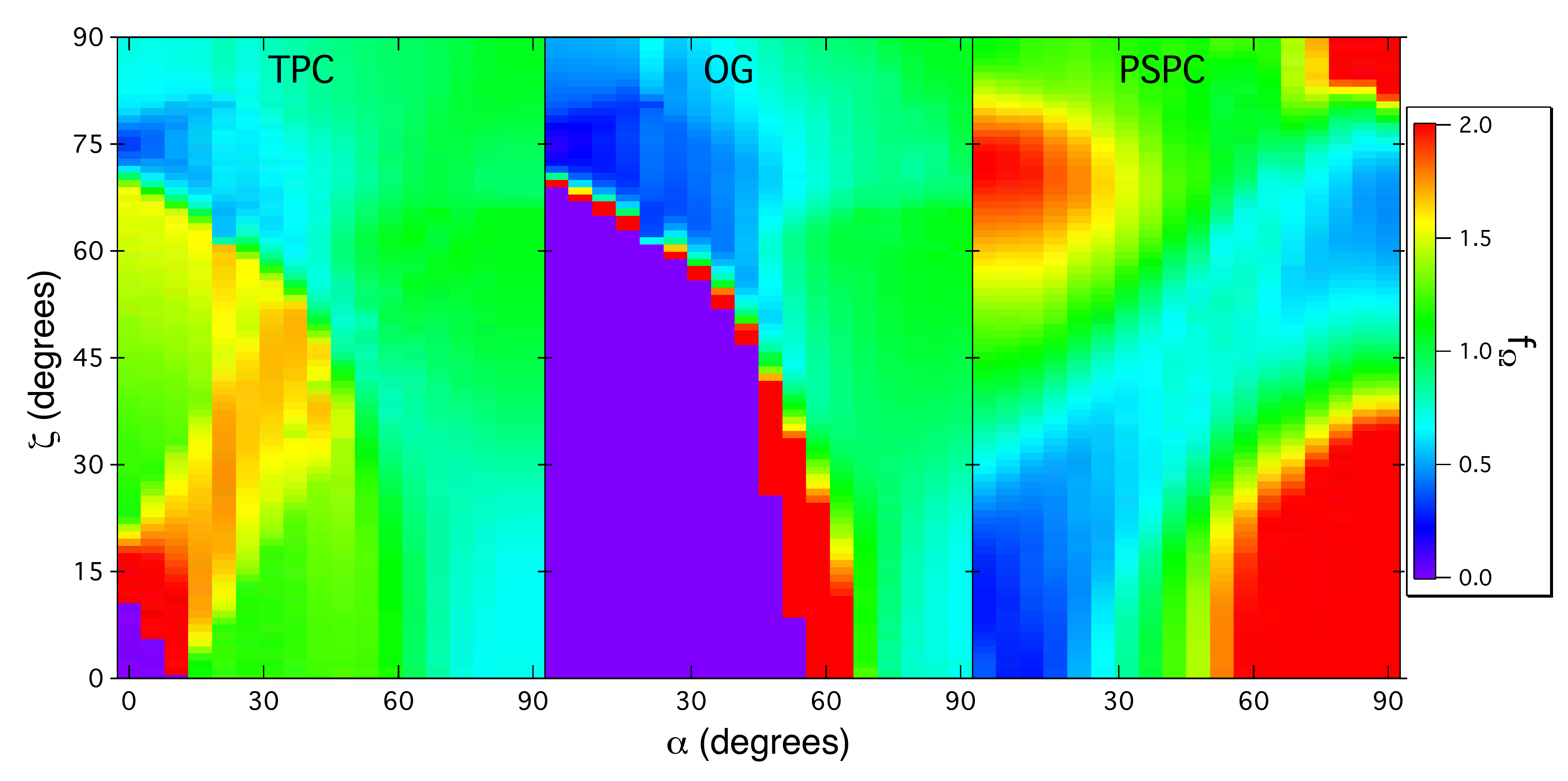}
\caption{The flux correction factor $f_\Omega$ plotted as a function of the viewing angle $\zeta$ versus the inclination angle $\alpha$. The values of $f_\Omega$ were obtained from the sky maps for the two-pole caustic (TPC), outer gap (OG) and pair starved polar cap (PSPC) for the $\gamma$-ray emission models that were used in fitting that radio and $\gamma$-ray light curves of the {\it Fermi} MSPs in the 2PC in the study of \cite{Johnson14} (see also \cite{Venter09} for more details).}
\label{fig:Foms}
\end{center}
\end{figure}

The $\gamma$-ray emission geometry is included in the sky maps of the $\gamma$-ray intensity patterns in viewing angle $\zeta$ vs. rotation phase angle $\phi$ as a function of the inclination angle used in the study of \cite{Johnson14} and \cite{ Venter09,Venter12} for three $\gamma$-ray models, the Outer Gap (OG), the Slot Gap Two-Pole Caustic (TPC), and the pair starved polar cap (PSPC) models.  Specifically the sky maps have been calculated for $P$= 5 ms with the electrons placed uniformly inside the annulus defined by the inner boundary from $\xi$ = 0.95 to outer value of 1.0 of the polar cap radius with the gap starting at 0.12 of the light cylinder ($R_{L C}$) and extending to 1.2 $R_{L C}$.   Further details associated with these sky maps can be found in \cite{Venter09}.   \cite{Johnson14} implemented these, and variants of these, emission models to fit the radio and $\gamma$-ray light curves of all of the {\it Fermi} MSPs in the 2PC.    Each of the high-energy emission models consists of eighteen sky maps beginning with an inclination angle $\alpha$ of 5 degrees followed with sky maps of additional 5 degree increments.  For each viewing angle $\zeta$, we obtain the phase averaged profile of the normalized sky maps used to calculate the simulated $\gamma$-ray flux.  For each simulated MSP at a given $\alpha$ and $\zeta$, a linear interpolation is performed within the array containing the averaged pulse profiles to obtain the averaged profile $\bar\Phi$ for the simulated MSP.  The phase averaged simulated $\gamma$-ray flux can be obtained via the expression
\eqs{\label{eq:GamFx}
F_\gamma\;=\;& \frac{L_\gamma\, \bar\Phi}{d^2} }
The contribution from the opposite pole at $180 - \zeta$ is also included for each viewing angle.
If we represent the sky map as $F\left(\zeta,\phi\right)$, then the averaged pulse profile is
\eqs{\label{eq:AvePro}
\bar\Phi\left(\zeta\right)\;=\;&\frac{\int\limits_0^{2\pi} F\left(\zeta,\phi\right) d\phi}{2\pi\int\limits_0^{2\pi}\int\limits_0^\pi F\left(\zeta,\phi\right)\sin\zeta d\zeta d\phi} }
Comparing this phase averaged profile $\bar\Phi\left(\zeta\right)$ with Eq. (4) of \cite{Watters09}, we find the relationship between $\Phi\left(\zeta\right)$ and the beam correction factor $f_\Omega$ to be given by 
\eqs{\label{eq:AveFO}
f_\Omega \; = \; & \frac{1}{4 \pi \, \bar\Phi\left(\zeta\right)} }
The value of $f_\Omega(\zeta,\alpha)$ for each of the three $\gamma$-ray emission models are shown in Figure \ref{fig:Foms} with the color scale representing values from 0 to 2 defined via average pulse profiles that have nonzero values.  While there are large regions within the $\zeta$ and $\alpha$ space where $f_\Omega=1$, there are significant regions in which $f_\Omega$ is either above or below 1.  While the OG and TPC models have somewhat similar regions with similar values of $f_\Omega$, for example, for $\alpha>30^{\rm o}$, the geometry of the PSPC is very different.  The large purple region for the OG model corresponds to a phase average of zero where there is no emission taking place.

\section{Radio and gamma-ray detection} \label{sec:det}
Accepting a simulated MSP as being detected by a given radio survey requires that it first be in the region of the sky that was observed by the survey.  In addition, its phase averaged flux at the frequency of the survey must be larger than the $S_{\rm min}$ of that survey given by the standard Dewey formula \citep{Dewey85} having the form
\eqs{\label{eq:Dewey}
S_{\rm min}\; = \; & \frac{C_{\rm thres}\left[T_{\rm rec}+T_{\rm sky}\left(l,b\right)\right]}{G\sqrt{N_p\,B\,t}}\sqrt{\frac{W}{P-W}} }
where $C_{\rm thres}$ is the detection threshold or signal-to-noise (S/N), $T_{\rm rec}$ is the receiver temperature (K), $T_{\rm sky}(l,b)$ is the sky temperature obtained from the all-sky 408 MHz survey of \cite{Haslam82}, $G\ \left({\rm K\; Jy^{-1}}\right)$ is the telescope gain, $N_p$ is the number of polarizations (usually 2), $B$ (MHz) is the total band width, $t$(s) is the integration time, $P$ (ms) is the pulsar period, and $W$(ms) is the effective pulse width.  The effective pulse width $W$ is given in terms of various other time smearing effects by the expression
\eqs{\label{eq:Width}
W^2 \;=\;& W^2_{\rm o}+\tau^2+\tau_{\rm DM}^2+\tau_{\rm scat}^2+\tau^2_{\rm trial_{DM}} }
where $W_{\rm o}$ is the intrinsic pulse width assumed to be $0.05\,P$ (ms) (a value typically used in the Dewey formula for radio surveys), $\tau$ is the low pass filter time constant applied before sampling, or if unknown, an assumed value of twice the sampling time $\tau_{\rm sample}$ is used, $\tau_{\rm DM}$ is the dispersion smearing time over the frequency interval, $\Delta\nu$ (MHz), $\tau_{\rm scat}$ is the time broadening of the pulse due to interstellar scattering, and $\tau_{\rm trial_{DM}}$ is an additional smearing when sampling is performed at a DM different from the actual DM of the pulsar and corresponds to the fourth term in Eq. (2) for $W$ in the study by \cite{Dewey85}.  These pulse broadening terms are routinely calculated in our population code (see \cite{Gont02} for further details).

The ``detection" of a $\gamma$-ray pulsar within the code consists of comparing the simulated $\gamma$-ray energy flux of the pulsar to the \fermi point source threshold all-sky map in the 2PC (for details see section 8.2 and Figure 16 in \cite{Abdo2PC}; private communication, Toby Burnett).   The threshold map is in energy flux units and has been constructed assuming a typical pulsar energy cutoff $E_{\rm cut}\,=\,2$ GeV and spectral index $\Gamma\,=\,1.8$ \citep{Abdo2PC}. Given the viewing angle $\zeta$ and the inclination angle $\alpha$, the high-energy emission model provides the flux correction factor $f_\Omega$ with which the $\gamma$-ray luminosity and the pulsar distance allow one to obtain the average $\gamma$-ray energy flux that can be compared directly to the threshold map without requiring the $E_{\rm cut}$ and $\Gamma$ of the simulated pulsar spectrum.   In addition, since most \fermi MSPs required pointed radio observations to first discover their pulsations, excluding one discovered by a \fermi blind search, we imposed a restriction that the simulated MSPs not only be detected by \fermi but also have a phase averaged radio flux above $30\ \mu$Jy and a larger dispersion measure than 2.5 $\rm pc/cm^3$.  The radio detection of a pulsar typically requires a significant DM signal.  The radio signals from nearby pulsars tend to have a larger contribution of ``red" noise, therefore we impose a lower limit on the simulated DM of \fermi MSPs.  Within the code, the ``detection" of a radio MSP requires that it be located within the field of view of the survey and have a phase averaged flux above the $S_{\rm min}$ of the given survey at its observing frequency.  On the other hand, the ``detection" of a \fermi MSP requires that its average $\gamma$-ray flux be above the point source threshold at its location in the sky, as well as requiring that the MSP satisfies the minimum phase averaged radio flux and DM.  Since the identification of the type of pulsar is done separately within the code, a simulated MSP can be a \fermi MSP, a radio-loud MSP, or both.   In addition,  there is a group of radio-weak $\gamma$-ray MSPs that have averaged $\gamma$-ray fluxes above the point source threshold map and therefore, seen by \fermi as point sources; yet, they have radio fluxes below our imposed limit of $30\ \mu$Jy or have DMs below 2.5 $\rm pc/cm^3$.  We can then predict that with deeper radio observations beyond the sensitivities of those attempted at the present, these radio-weak \fermi MSPs can be discoverable as discussed later in Section \ref{sec:Proj}.

\section{Markov Chain Monte Carlo} \label{sec:MCMC}
A Markov Chain Monte Carlo (MCMC) is implemented here to explore a large portion of the parameter space of the set of free model parameters described in Table \ref{tab:Fpara}, which are the overall normalization factors and the exponents of the period and period derivative in the radio and $\gamma$-ray luminosity models given in Equations (\ref{eq:LumT}) and (\ref{eq:GamLum}).  Each random step in the chain determines the trial values of the free parameters, for which the code performs a Monte Carlo (MC) simulation that results in a set of 1-D histograms of the MSP features to compare to those observed.  In the case of exploring the free parameters associated with radio luminosity, we compare four 1-D histograms of the period $P$, period derivative $\dot P$, dispersion measure (DM), and radio flux $S_{1400}$ at 1400 MHz, whereas in the case of the $\gamma$-ray luminosity, we compare the same four and an additional 1-D histogram of the $\gamma$-ray energy flux (above 100 MeV).  As discussed at the end of Section \ref{sec:det}, although within the code the ``detection" of radio and \fermi MSPs is separate, it is still necessary to search a 6-D parameter space for the whole set of free parameters given that the uncertainties in radio parameters influence the $\gamma$-ray parameters and vice versa.  

\begin{deluxetable}{cl}[ht!]
\tablecolumns{2}
\tablewidth{0pc}
\tablecaption{Set of free parameters used in the MCMC\label{tab:Fpara} }
\tablehead{
\colhead{Parameter} & \colhead{Description} }
\startdata
$f_\nu$ & Overall multiplicative factor of the radio luminosity \\
$\alpha_\nu$ & Period exponent of the radio luminosity \\
$\beta_\nu$ & Period derivative exponent of the radio luminosity \\
$f_\gamma$ & Overall multiplicative factor of the $\gamma$-ray luminosity \\
$\alpha_\gamma$ & Period exponent of the $\gamma$-ray luminosity \\
$\beta_\gamma$ & Period derivative exponent of the $\gamma$-ray luminosity \\
\enddata 
\end{deluxetable}

Each chain of the MCMC, using a different initial random seed, begins in a different place in the 6-D space determined by a uniform distribution of the step size for each of the parameters.  
At each step, the MC simulation is run with the new set of the free parameters, using as input the group of MSPs spatially evolved to the present time as described in Section \ref{sec:evolved}.   Each simulated MSP is given a randomly selected magnetic field from Equation (\ref{eq:Bf}), a mass accretion birth line selected from Equation (\ref{eq:Po}), random viewing and inclination angles, and the present day period  and period derivative from Equation (\ref{eq:PresP}) and Equation (\ref{eq:Spin}).  The free model parameters do not appear in closed form in the simulated histograms as there is no direct analytic description of the probability distributions of the resulting 1-D histograms.  
The radio luminosity is determined from Equation (\ref{eq:LumT}), and the core and cone beam luminosities are assigned with  Equations (\ref{eq:LumRC}) and (\ref{eq:CrCn}).  A pulse profile is calculated for each simulated MSP to obtain the averaged radio flux.  If the effective width of the profile is larger than 60\%, the MSP is not considered a radio pulsar.  The location of the MSP in the sky determines which radio surveys to consider.  The survey flux threshold $S_{\rm min}$ obtained from the Dewey formula given in Equation (\ref{eq:Dewey}) is compared to that of the simulated MSP to determine its detectability.  Regardless of whether the simulated MSP is identified as a ``detected" radio MSP, the $\gamma$-ray characteristics are considered separately, the 
$\gamma$-ray luminosity is obtained via Equation (\ref{eq:GamLum}), the phase averaged profile intensity is obtained from the high-energy emission model sky maps, and the $\gamma$-ray energy flux is obtained with Equation (\ref{eq:GamFx}).  If the $\gamma$-ray flux is larger than the \fermi point source threshold for the MSP's given location in the sky, the MSP is selected as a \fermi MSP.  If the conditions are satisfied, the simulated MSP can be identified as a {\it Fermi}-detected MSP.  If this is not the case, the information of the simulated MSP is written to a computer file and analyzed as MSPs that contribute to the $\gamma$-ray diffuse background.

The selected characteristics of the simulated MSPs from the MC at each step are binned into histograms to obtain the log-likelihoods as follows.
The likelihood function is constructed to take into account the Poisson statistics of the small number of pulsars $\cal N$ in the bins of the selected histograms, given by the expression \citep{Hauschild01, Bevington03, James06}:
\eqs{     {\cal L}_j\;  = \; &   \prod_i^{N_j}\frac{  e^{ -{\cal N}^{sim}_i }\left[\ {\cal N}^{sim}_i\ \right]^{{\cal N}^{det}_i}} { \left( {\cal N}^{det}_i\right)! } ,}
for a given histogram of $N_j$ bins, where ${\cal N}^{det}_i$ and ${\cal N}^{sim}_i$ represent the number of counts in bin $i$ of the histograms of the detected and simulated pulsars, respectively.  The log-likelihood $\ln {\cal L}_j$ for each pulsar characteristic being compared are summed together to provide the total log-likelihood ${\cal L}_{\rm tot}$ for each random step taken. 
\eqs{ \ln {\cal L}_{\rm tot} \;=\; \sum\limits_{j=1}^{N_h}\, \ln{\cal L}_j,}
where $N_h$ is the number of 1-D histograms being compared and where $j$ is referring to the histograms. 
A total of nine histograms are used for radio ($N_h=4$) and \fermi ($N_h=5$) MSPs.  The distributions are binned in histograms with 10 bins.  To obtain smoother simulated histograms, the evolved group consists of ten times the number required by the birthrate, then we scale the histograms accordingly.  
As prescribed with a typical Metropolis-Hasting algorithm \citep{Hastings70,Liang10} if the log-likelihood $\ln{\cal L}$ increases, the step in the 6-D space is accepted.  The step is also accepted if a random number is less than $e^{\Delta\ln{\cal L}}$, where $\Delta\ln{\cal L}$ is the difference between the current likelihood  and the previous likelihood. Otherwise, a new step is initiated from the previously accepted location.  
  In the MCMC simulation the accepted steps occur only after a large number of steps are taken.  Therefore, there is no need to thin the accepted steps. The information of each of the accepted steps is written to a file.
A fairly large number of steps in each chain are taken before a step is accepted.  As a result, the accepted steps are not sequential.  
The accepted steps of the MCMC simulations represent draws from the posterior probability distribution of the set of the six free parameters for the radio and $\gamma$-ray luminosity models. Once the values of the model parameters that maximize the likelihood are determined, a separate MC code is used to simulate MSPs.  The set of accepted MCMC steps for each high-energy emission model is later used to propagate the uncertainties of the model parameters to obtain the uncertainties of the results.  Computer files are  generated that include the characteristics of radio and \fermi detected MSPs as well as those that are not detected, contributing to the $\gamma$-ray diffuse background for later analysis.

\section{Results}

\subsection{Markov Chain Monte Carlo}
In the MCMC code the steps of the free model parameters are randomly selected assuming a uniform distribution determined by a constant maximum step size.  The parameter space is widely explored by the chains without any assumed probability distribution of the parameter space.  As expected given the nature of the MC simulation and the limited number of detected MSPs with which to compare, the likelihood fluctuates about a particular value after the burn-in period.
All the accepted steps after the burn-in period of all the chains are gathered.  

\begin{figure}[ht!]
\begin{center}
\includegraphics[trim=0in 0in 0in 0in, width=7.5in]{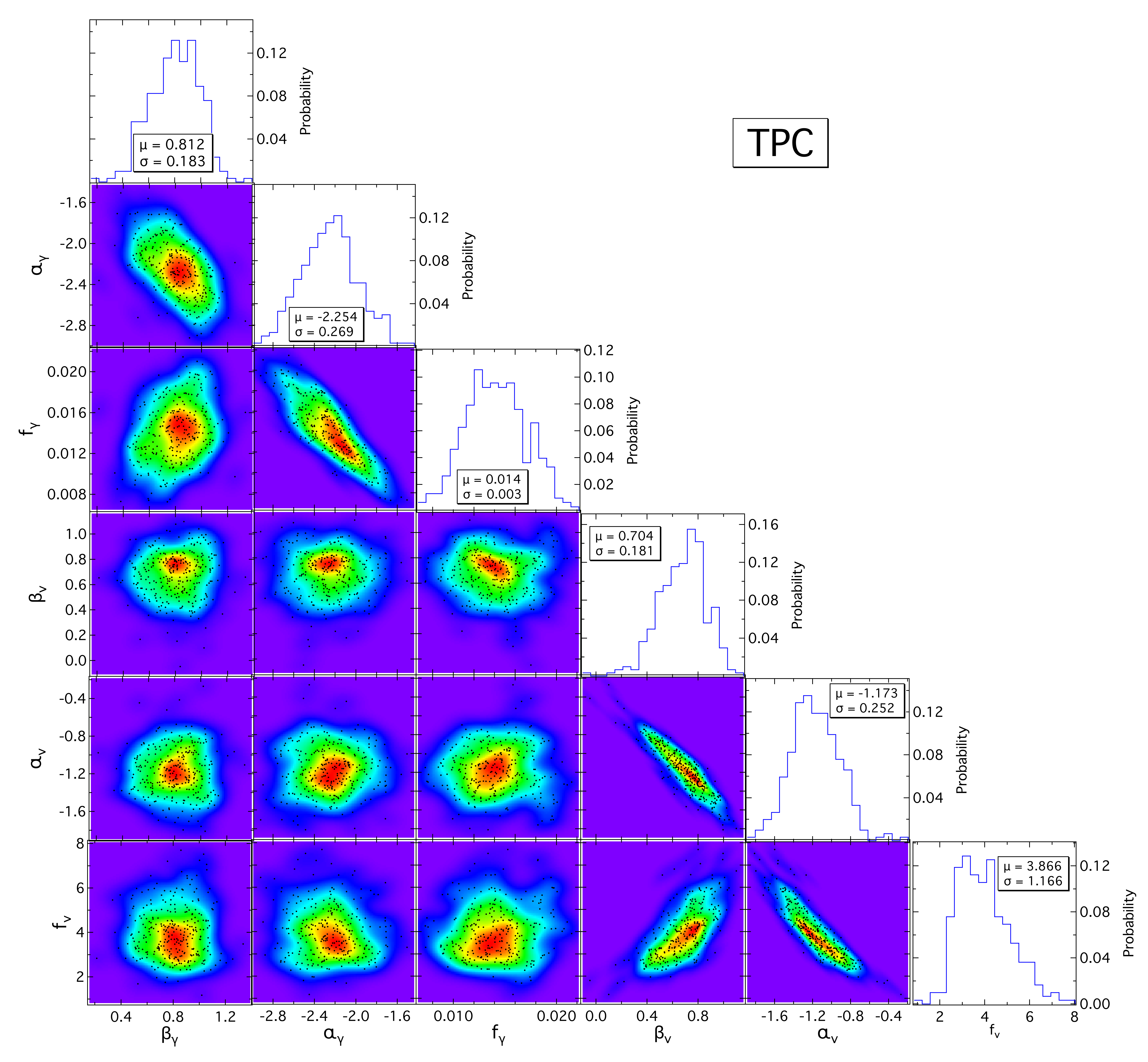}
\caption{A corner plot for the MCMC results for the case of the high-energy model TPC, representing the 2-D marginal distributions of all combinations of the six free model parameters plotted against each other as well as the 1-D marginal distributions of each of the free parameters.  The color of the image represents the 2-D marginal probability distribution generated with a Kernel Density Estimator \citep{Scott15} using 2-D Gaussian kernels.  The black dots in the 2-D image plots are the accepted steps in the MCMC after a reasonable burn in period.  The median and standard deviations of the 1-D distributions are indicated in the boxes of the plots.  The Pearson's correlation coefficients for the free parameters are presented in Table \ref{tab:Pearson} for the correlations of all possible pairs of free parameters of the $\gamma$-ray luminosity with those of the radio luminosity.  The red numbers suggest a lack of strong correlation between the luminosities. } 
\label{fig:MCMCa}
\end{center}
\end{figure}

In Figure (\ref{fig:MCMCa}), we display the MCMC results as 2-D and 1-D marginalized distributions in a corner plot for the TPC high-energy model.  All possible pairs of the six model parameters are displayed as 2-D image plots via the Kernel Density Estimation  \citep{Scott15}  using a 2-D Gaussian kernel with the {\it ks} package \citep{Duong07} of the R software package for \citep{Rproj} on each of the accepted MCMC steps after a burn-in period.  The image plots of 
\noindent pairs of parameters pertaining to the same luminosity model display significant correlations especially for those of the radio luminosity model with the pair $f_\gamma\ {\rm vs.}\ \beta_\gamma$ plot showing the least correlation between these pairs.  The image plots of the pairs of parameters from different luminosity models are almost circular patterns indicating little correlation between the pairs, suggesting that the radio and $\gamma$-ray luminosity models are not strongly correlated and that their parameters show little correlation with each other.  The 1-D marginal distributions are also included within the corner 
\begin{figure}[ht!]
\begin{center}
\includegraphics[trim=0in 0in 0in 0in, width=7.5in]{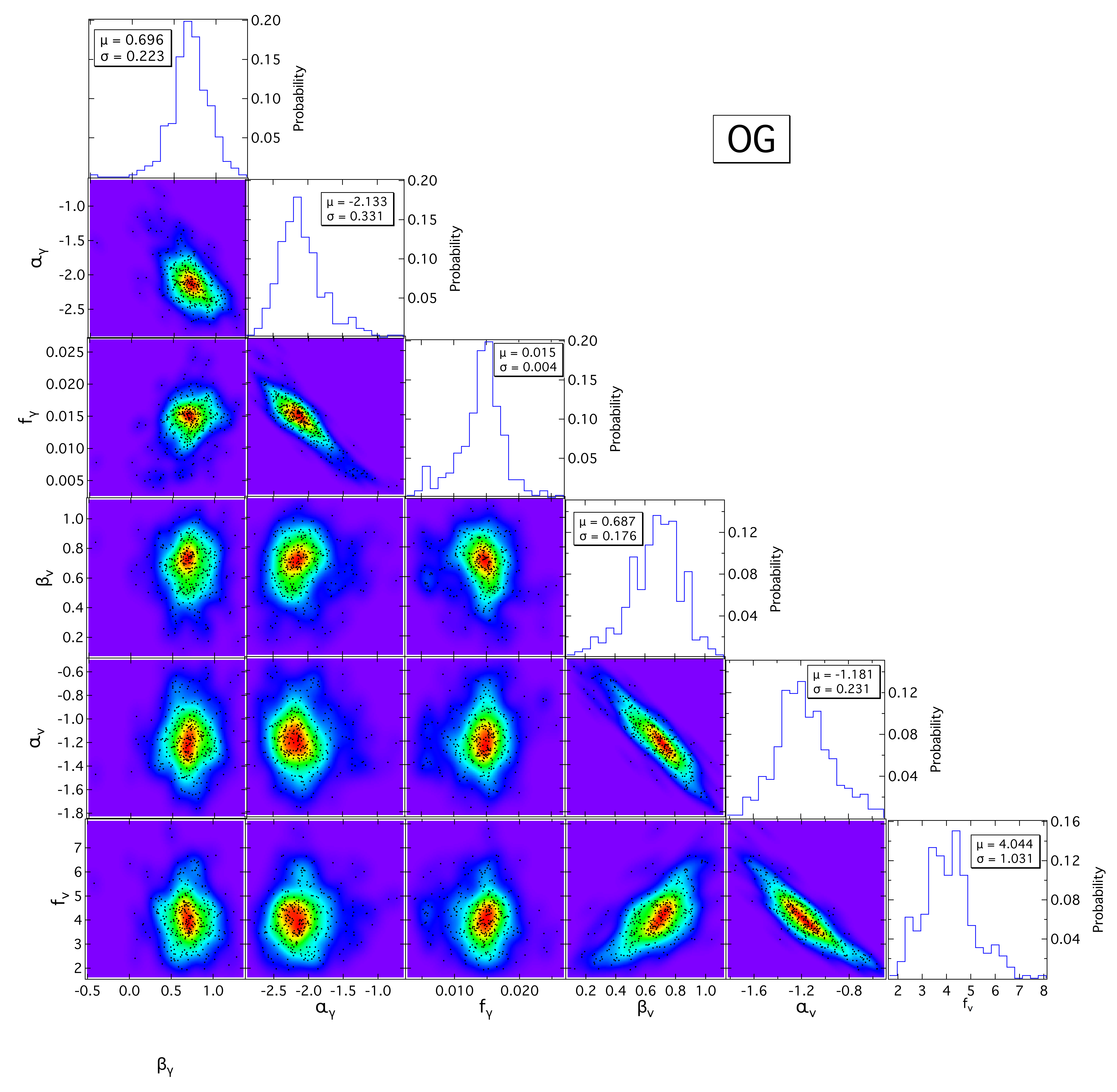}
\caption{A corner plot for the MCMC results for the case of the high-energy model OG, representing the 2-D marginal distributions of all combinations of the six free model parameters plotted against each other as well as the 1-D marginal distributions of each of the free parameters.  See Figure \ref{fig:MCMCa}. } 
\label{fig:MCMCb}
\end{center}
\end{figure}
\noindent plot and give us some idea of the shape of the projection of the 6-D distribution onto the dimension of each of the single parameters.  We also indicate in the boxes of the graphs of each of the 1-D distributions  the median and standard deviation of the distributions, providing some idea of the uncertainties of the parameters of the luminosity models.  Given that these are marginalized distributions, the widths are not precisely the uncertainties of the parameters, but do provide some idea of their values.  We show similar information of the MCMC results for the OG and PSPC 
high-energy models in Figures (\ref{fig:MCMCb}) and (\ref{fig:MCMCc}), respectively.  To provide some general idea of the statistical correlations between the model parameters, we provide Pearson's correlations coefficients between all pairs of the \begin{figure}[ht!]
\begin{center}
\includegraphics[trim=0in 0in 0in 0in, width=7.5in]{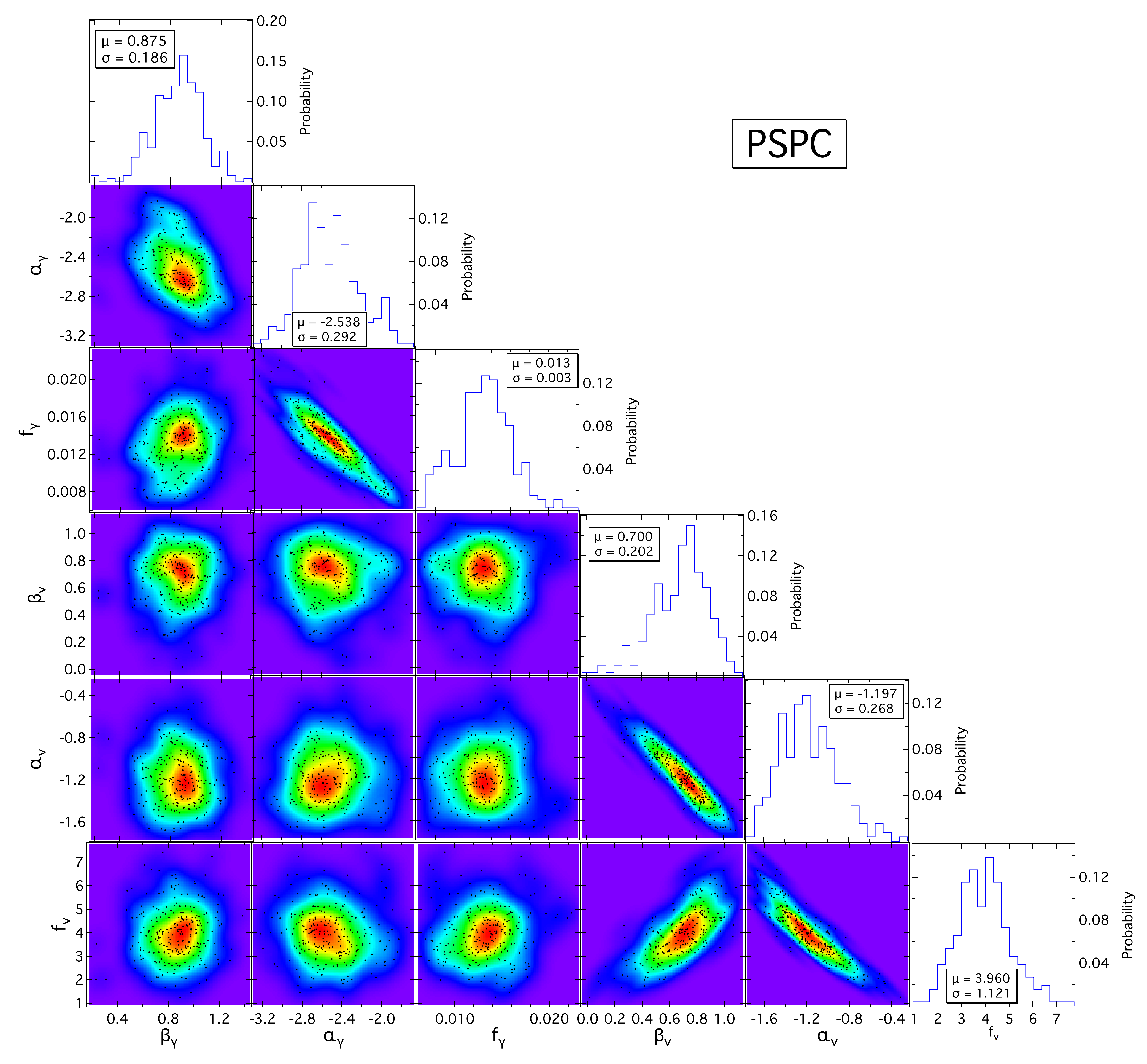}
\caption{A corner plot for the MCMC results for the case of the high-energy model PSPC, representing the 2-D marginal distributions of all combinations of the six free model parameters plotted against each other as well as the 1-D marginal distributions of each of the free parameters.  See Figure \ref{fig:MCMCa}. } 
\label{fig:MCMCc}
\end{center}
\end{figure}
parameters for each of the models in Table \ref{tab:Pearson}.  A small negative or positive coefficient suggests that that pair of model parameters have a small and insignificant correlation between them.  The group of coefficients in red are the smallest in absolute value and represent those correlations between the possible pairs of parameters of the radio and $\gamma$-ray luminosity models, suggesting the statistical independence of these luminosity models.  This is indeed expected as, for the most part, the selection criteria where a radio or $\gamma$-ray MSP is ``detected" within the simulation are separate with the exception that the pulsations of the \fermi MSPs were actually discovered in dedicated deep pointed radio observations.  Thus we require some radio flux to be associated with simulated \fermi radio-loud MSPs (see section \ref{sec:selgrp}).  This requirement then motivates the 6-D MCMC exploration of the parameter space of the two models.  Though separate 3-D MCMC explorations do yield similar results, the 6-D exploration provides a more statistically rigorous framework for discussion.  The $\alpha$ parameter associated with the exponent of the $P$ in the models indicates the strongest correlations with the overall factors $f$ of the luminosity models. The next strongest correlations are between the $P$ and $\dot P$ exponents $\alpha$ and $\beta$ in both models and in all three high-energy models.  While one might find subtle differences among the three high-energy models, the 2-D image and 1-D distributions as well as the statistics do not suggest significant differences.
\begin{deluxetable}{c|c|c|c|c|c|c}[!ht]
\tablecolumns{7}
\tablewidth{0pc}
\tablecaption{MCMC results: Pearson's Correlation Coefficients \label{tab:Pearson} }
\tablehead{
Parameter 	&	$ f_\nu $ & $ \alpha_\nu $ & $ \beta_\nu $ & $ f_\gamma $ & $ \alpha_\gamma $ & $ \beta_\gamma $ }
\startdata
 \multicolumn{7}{c}{Two Pole Caustic High-Energy Model } \\ \hline
 $ f_\nu $ 			& 1.0				&   				&					&			& 		& \\
$ \alpha_\nu $ 		& -0.889			& 1.0				&					& 			& 		&\\
$ \beta_\nu $ 		& 0.648			&-0.906			& 1.0					& 			& 		&  \\
$ f_\gamma $ 		& {\color{red} 0.082}	& {\color{red} -0.013}	& {\color{red} -0.028}		& 1.0			&		&   \\
$ \alpha_\gamma $ 	& {\color{red} -0.124}	&  {\color{red} 0.065}& {\color{red} -0.033}		& -0.841		& 1.0 	&  \\
$ \beta_\gamma $	& {\color{red} 0.007}	& {\color{red} -0.013}	& {\color{red}  0.036} 	& 0.293		& -0.542	& 1.0 \\
\hline
\multicolumn{7}{c}{Outer Gap High-Energy Model } \\ \hline
$ f_\nu $ 			& 1.0				&   				&					&			& 		& \\
$ \alpha_\nu $ 		& -0.875			& 1.0				&					& 			& 		&\\
$ \beta_\nu $ 		& 0.606			&-0.891			& 1.0					& 			& 		&  \\
$ f_\gamma $ 		& {\color{red} -0.078}	& {\color{red} 0.110}	& {\color{red} -0.120}		& 1.0			&		&   \\
$ \alpha_\gamma $ 	& {\color{red} 0.076}	&  {\color{red} -0.076}& {\color{red} 0.062}		& -0.871		& 1.0 	&  \\
$ \beta_\gamma $	& {\color{red} -0.082}	& {\color{red} 0.062}	& {\color{red}  -0.034} 	& 0.325		& -0.524	& 1.0 \\
\hline
\multicolumn{7}{c}{Pair Starved Polar Cap High-Energy Model } \\ \hline
$ f_\nu $ 			& 1.0				&   				&					&			& 		& \\
$ \alpha_\nu $ 		& -0.889			& 1.0				&					& 			& 		&\\
$ \beta_\nu $ 		& 0.678			&-0.924			& 1.0					& 			& 		&  \\
$ f_\gamma $ 		& {\color{red} 0.047}	& {\color{red} -0.040}	& {\color{red} 0.025}		& 1.0			&		&   \\
$ \alpha_\gamma $ 	& {\color{red} -0.047}	&  {\color{red} 0.045}& {\color{red} -0.047}		& -0.854		& 1.0 	&  \\
$ \beta_\gamma $	& {\color{red} 0.066}	& {\color{red} -0.052}	& {\color{red}  0.043} 	& 0.165		& -0.395	& 1.0 \\
\enddata
\end{deluxetable}

\begin{figure}[ht!]
\begin{center}
\centerline{\includegraphics[trim=0in 0in 0in 0in, width=4.5in]{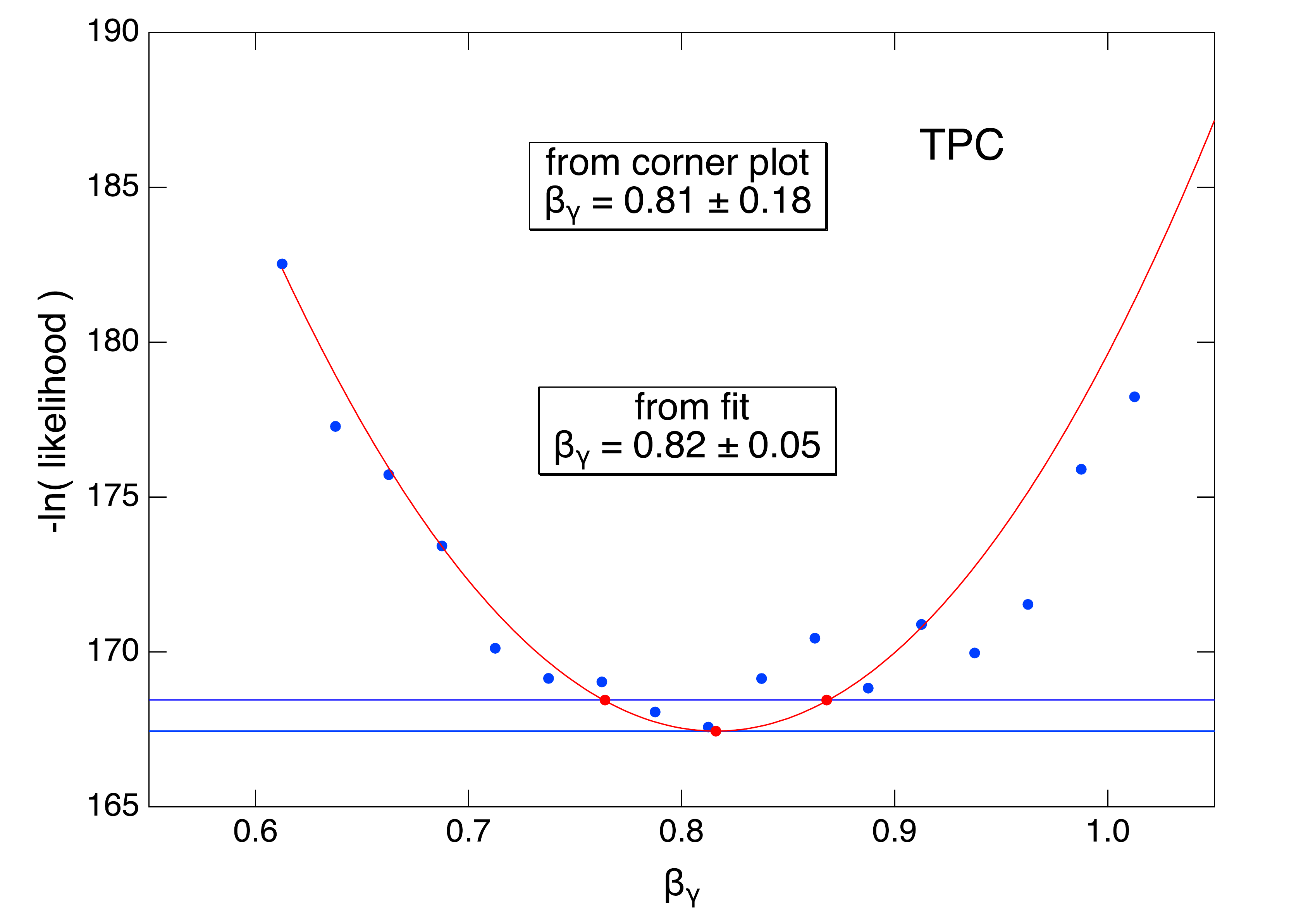} }
\caption{An example of the distribution of binning the maximum likelihoods of the accepted MCMC steps after a burn in period as a function of the free parameter $\beta_\gamma$, which is the exponent of the $\dot P$ dependence of the $\gamma$-ray luminosity.  Due to the MC nature of each simulated step, there is a ``random noise" associated with each of the accepted MCMC steps.  As a result, the maximum likelihood values for each bin of $\beta_\gamma$ do not follow a continuous smooth curve, but fluctuate.  To find the overall maximum likelihood occurrence of $\beta_\gamma$, we fit the distribution with a quadratic.  The results of the maximum likelihood value and the 1 $\sigma$ drop from the maximum likelihood value for 1 degree of freedom are shown in the box and can be compared to the result of the overall marginalized distribution of accepted values as a function of $\beta_\gamma$ shown previously in the corner plot for the TPC case in Figure \ref{fig:MCMCa}.  The full results are presented in Table \ref{tab:paraunc}.} 
\label{fig:MaxBeta}
\end{center}
\end{figure}

In order to obtain the uncertainties of the model parameters, we perform separate MCMC simulations in which the values of the model parameters are assumed to be normally distributed with means and widths set at the medians and standard deviations of the 1-D distributions indicated in the corner plots for each of the high-energy models displayed in Figures (\ref{fig:MCMCa}), (\ref{fig:MCMCb}), and (\ref{fig:MCMCc}).  For these MCMC simulations, we must perform additional MC simulations near the maximum likelihood values of the parameter space.  We perform 3000 MCMC steps for each of the high-energy models, accepting all of the steps.
We then bin the MCMC steps in 1-D histograms for each parameter selecting the maximum likelihood of the step falling within the bin.
As an example, we show in Figure \ref{fig:MaxBeta} the binned results of maximum likelihood (solid blue circles) for the $\beta_\gamma$ parameter of the $\gamma$-ray luminosity model within the TPC high-energy model.  Given the nature of the MC simulation, each step represents an entirely different group of random numbers which generate the simulated histograms to compare to those observed.  As a result, the maximum likelihood tends to fluctuate and does not follow a well-defined continuous function.  To find the overall maximum likelihood, we perform 
a quadratic fit without implying any prior distribution shown in the figure as a red curve with the maximum and the drop of 1 in the negative log-likelihood $-\ln{\cal L}$ represented as solid red circles.  The drop of 1 from the maximum likelihood represents $\Delta\chi^2 = 1$ for 1 degree-of-freedom as suggested by Wilk's theorem \citep{wilks38}, representing the 1 $\sigma$ uncertainty of the parameter.   Implementing this method, we obtain the uncertainties of the six parameters of the luminosity models for each of the high-energy models that are included in Table \ref{tab:paraunc}.  The estimated most likely values of the free parameters and their uncertainties are expected all to fall within the medians and standard deviations of the 1-D distributions in the corner plots, though the estimated uncertainties are smaller than the standard deviations of the 1-D marginalized distributions, which are projections of the 6-D distribution unto the 1-D parameter space.
\begin{deluxetable}{c|c|c|c}[!ht]
\tablecolumns{4}
\tablewidth{0pc}
\tablecaption{MCMC results: the estimated most likely values of the free parameters \label{tab:paraunc} }
\tablehead{
Free Parameter  &  Two Pole Caustic Slot Gap (TPC) & Outer Gap (OG) & Pair Starved Polar Cap (PSPC) }
\startdata
$f_\nu$			&      $3.67  \pm 0.36$	&	$4.06\pm 0.60$		&	$4.17\pm 0.58$		\\
$\alpha_\nu$ 		&	$-1.13 \pm 0.09$	& 	$-1.17\pm 0.13$ 	& 	$-1.19\pm 0.13$ 	\\
$\beta_\nu$	 	& 	$0.65  \pm 0.08$	&	$0.64 \pm 0.10$	&	$0.69 \pm 0.11$	\\
$f_\gamma$		&  $0.0122 \pm 0.0012$	&	$0.0116\pm 0.0022$	&	$0.0117\pm 0.0021$	\\
$\alpha_\gamma$ 	& 	$-2.12\pm 0.10$	&	$-1.93\pm 0.18$	& 	$-2.43\pm 0.17$	\\
$\beta_\gamma$ 	& 	$0.82 \pm	0.05$	&	$0.75 \pm 0.13$	& 	$0.90 \pm 0.12$	\\
\enddata
\end{deluxetable}

In order to test the results obtained with the MCMC code, we ran our MC code to generate synthetic data, assuming the TPC high-energy model with four sets of the model parameters $f_\nu$, $\alpha_\nu$, $\beta_\nu$, $f_\gamma$, $\alpha_\gamma$, and $\beta_\gamma$.  The MC code produced 92 radio pulsars and 54 \fermi pulsars.  Histograms of the pulsar characteristics were made and treated as the observed data in the MCMC simulations.  We ran the MCMC code, searching the 6-D parameter space associated with the luminosity models, and acquired accepted steps after a burn-in period.   We obtained marginalized 1-D distributions similar to those in the corner plots from the accepted steps to obtain an idea of how close the MCMC simulation was in obtaining the parameter values of the generated synthetic data.  We tabulate the results of four markedly different cases of synthetic data in Table \ref{tab:syndata} where the ``Set Value" represents the value of the parameter that was used to generate the synthetic data and the ``MCMC Dis." is the resulting median and standard deviation of the 1-D marginalized distributions from the MCMC simulation for each of these cases.  In all of these cases, the MCMC simulation does indeed focus on the parameter region of the set of parameters that were used to generate the synthetic data, with one exception in the case of the $f_\nu$ parameter for Case 2, which only lies slightly outside.
\begin{deluxetable}{c|c|c|c|c|c|c|c|c}[ht!]
\tablecolumns{9}
\tablewidth{0pc}
\tablecaption{Synthetic Data and MCMC results: \label{tab:syndata} }
\tablehead{   & \multicolumn{2}{c|}{Case 1} & \multicolumn{2}{c|}{Case 2} & \multicolumn{2}{c|}{Case 3} & \multicolumn{2}{c}{Case 4} \\ \hline
Free Parameter 	&	Set Value	&	MCMC Dis. &	Set Value	&	MCMC Dis.  &	Set Value	&	MCMC Dis. &	Set Value	&	MCMC Dis. }
\startdata
$f_\nu$			&      $6.2$	&	$7.0    \pm  1.6$	&      $3.3$	&	$4.8    \pm  1.4$	&      $2.24$	&	$2.72    \pm  0.72$	&      $0.657$	&	$0.80    \pm  0.22$	\\
$\alpha_\nu$ 		&	$-1.6$	&	$-1.39 \pm  0.22$	&	$-1.0$	&	$-1.07 \pm  0.23$	&	$-1.5$	&	$-1.62 \pm  0.18$	&	$-0.5$	&	$-0.60 \pm  0.19$	\\
$\beta_\nu$	 	& 	$0.5$	&	$0.43  \pm  0.18$	& 	$0.2$	&	$0.24  \pm  0.16$	& 	$1.3$	&	$1.34  \pm  0.12$	& 	$0.7$	&	$0.74  \pm  0.15$	\\
$f_\gamma$		&  	$0.016$	&	$0.019 \pm 0.008$	&  	$0.006$	&	$0.0083 \pm 0.0030$&  	$0.00676$&	$0.0085 \pm 0.0034$&  	$0.00238$&	$0.0029 \pm 0.0008$\\
$\alpha_\gamma$ 	& 	$-2.5$	&	$-2.6    \pm 0.50$	& 	$-1.5$	&	$-1.6    \pm 0.37$	& 	$-2.0$	&	$-2.21    \pm 0.43$	& 	$-1.0$	&	$-1.14    \pm 0.26$	\\
$\beta_\gamma$ 	& 	$0.4$	&	$0.40   \pm 0.22$	& 	$0.0$	&	$0.17   \pm 0.25$	& 	$1.3$	&	$1.33   \pm 0.18$	& 	$1.0$	&	$1.05   \pm 0.15$	\\
\enddata
\end{deluxetable}

\subsection{Detected and simulated distributions}
We conclude that the most likely values of the free parameters of the luminosity models from the MCMC simulations in Table \ref{tab:paraunc} provide us with the adequate set of parameters to use in our final MC simulations.  However, the reason for an MC approach is that there is no analytic description of the simulation, therefore it is not possible to propagate the uncertainties of the parameters in Table \ref{tab:paraunc}.  Instead, we use the MCMC accepted steps to propagate the uncertainties as they sample the posterior 6-D distribution of the parameter space.  We assume a fixed birth rate of 4.5 MSPs per Myr (SGH) and process an evolved group of MSPs with ten times as many MSPs as required by the birth rate for each of over 300 accepted steps for each high-energy model.  \citep{Scott15}.

\begin{figure}[h!]
\begin{center}
\includegraphics[trim=0.5in 0in 0in 0in, width=5.in]{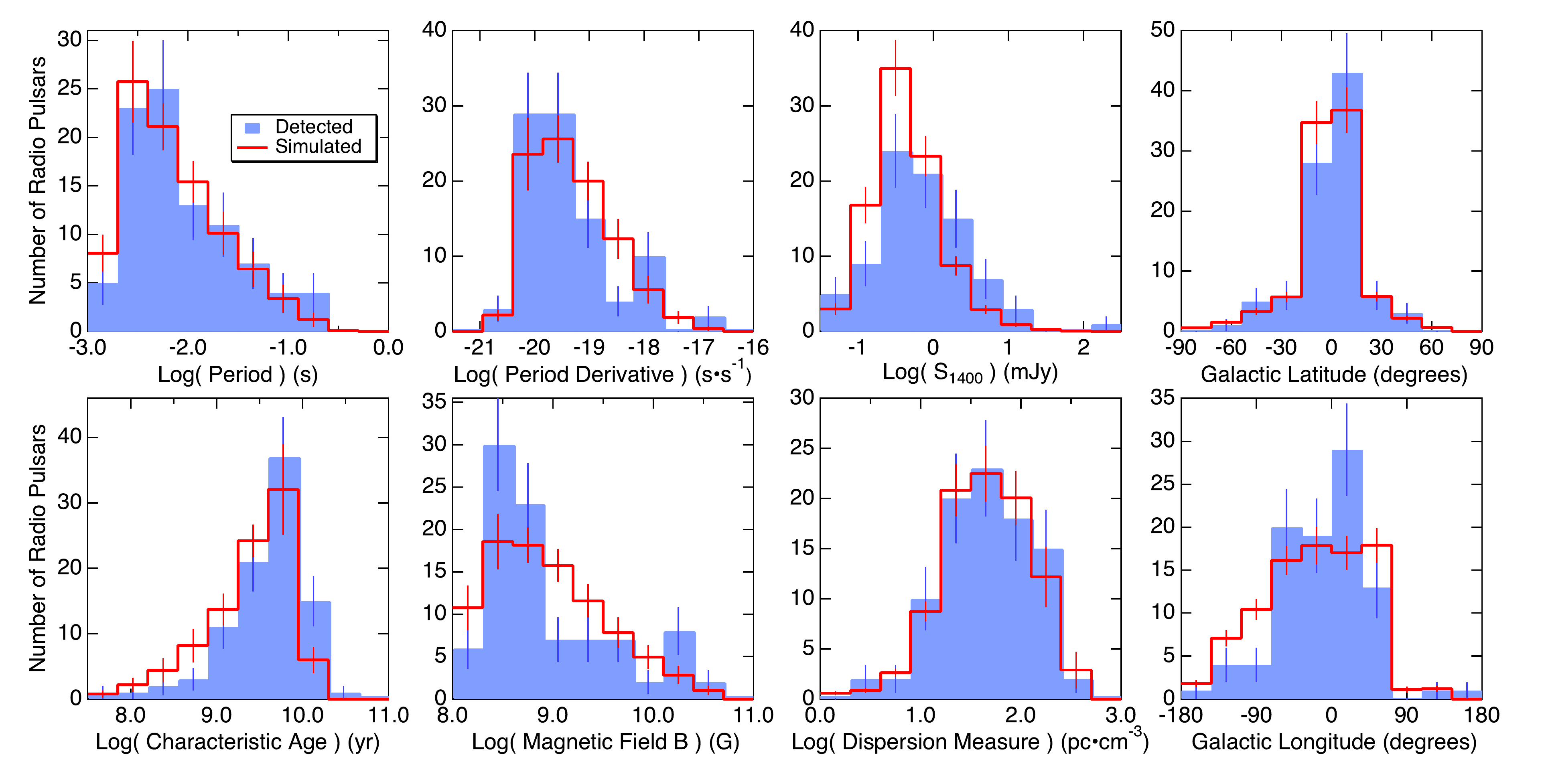}
\caption{Distributions of indicated radio pulsar characteristics for detected (solid blue histograms) and for simulated using the core-cone radio emission geometry (open red histograms).  The MC simulated histograms and their uncertainties represent the average and standard deviations of all the accepted steps after a burn-in period of the MCMC simulation.  The uncertainties of the detected histograms are Poisson.}
\label{fig:Radio}
\end{center}
\end{figure}

Various radio MSP characteristics are compared in Figure \ref{fig:Radio} for detected (solid blue histograms) and simulated (open red histograms) distributions.   Four of these characteristics, period, period derivative, dispersion measure (DM), and radio flux at 1400 MHz ($S_{1400}$), are directly measured, while the characteristic age and magnetic field are inferred qualities from the period and period derivative.  In order to obtain a simulated DM, the MC code implements the NE2001 electron density of the Galaxy \citep{CorLaz03}, which introduces significant uncertainty (approximately 20\%---40\%) associated with the simulated DM.   The histograms represent the averages and their uncertainties represent the standard deviations of the accepted steps of the MCMC simulations. The uncertainties of the detected histograms are Poisson, while those of the simulated bins are the standard deviations of the predictions of the accepted steps.  Since the simulation of radio MSPs in the code is separate from that of the $\gamma$-ray MSPs, the histograms of the radio MSP features are independent of the assumed high-energy model. The agreement of the MC simulation is acceptable for the histograms associated with the radio MSP  characteristics with the exception of some systematic differences, for example in histograms of the radio flux $S_{1400}$ with the simulation predicting more MSPs with smaller fluxes and in the predicted magnetic field $B$ distributions where more low field MSPs are observed than predicted.

\begin{figure}[h!]
\begin{center}
\includegraphics[trim=1in 0in 0in 0in, scale=0.5]{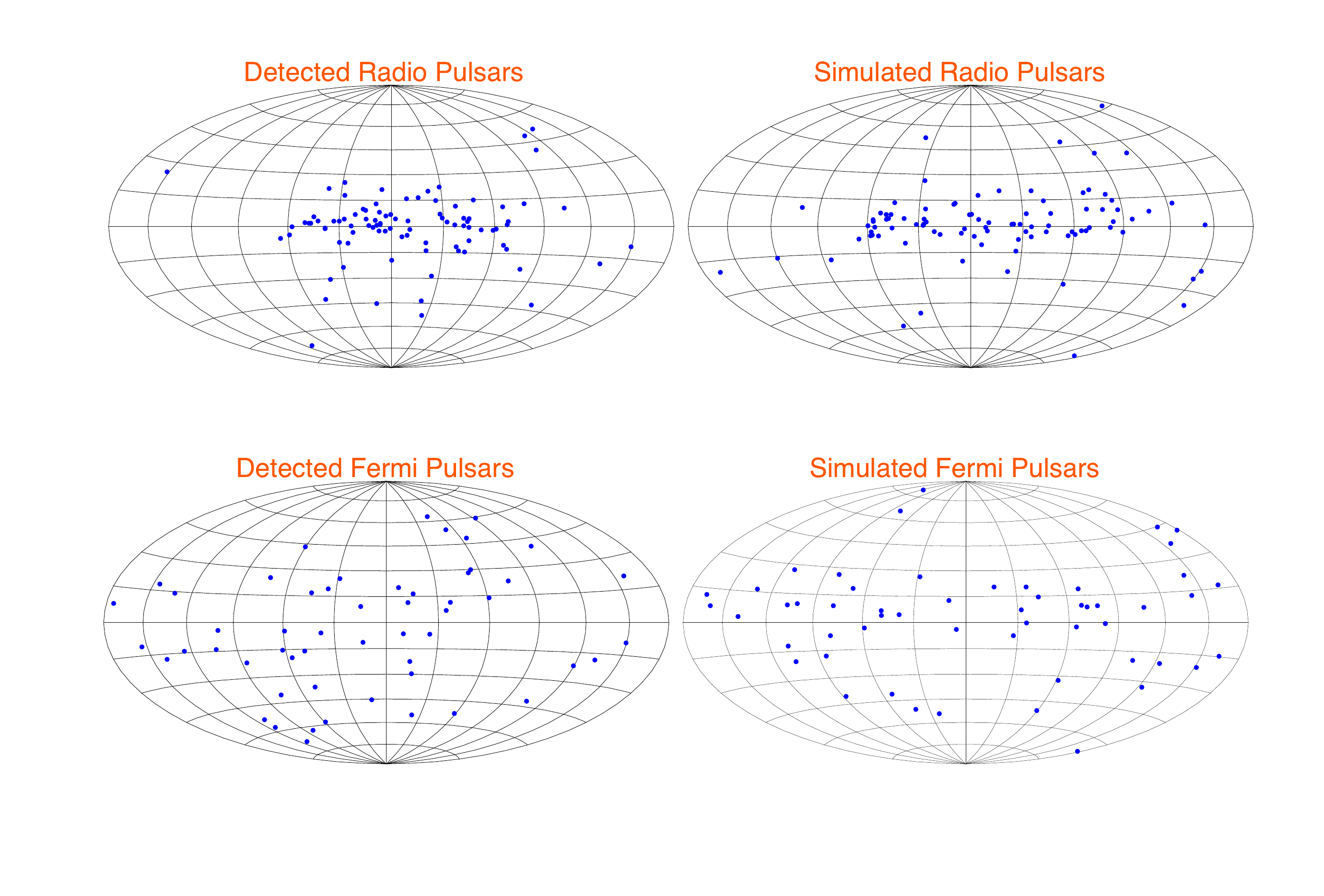}
\caption{Hammer-Aitoff projections of 92 radio and 54 \fermi pulsars detected (left panels) and simulated (right panels).  The simulated \fermi pulsars assume the TPC high-energy emission model.  The simulation was performed using the values of the parameters of the luminosity models in Table \ref{tab:paraunc} that maximize the corresponding likelihoods.}
\label{fig:Aitoff}
\end{center}
\end{figure}

\begin{figure}[h!]
\begin{center}
\includegraphics[trim=0in 0in 0in 0in, width=4in]{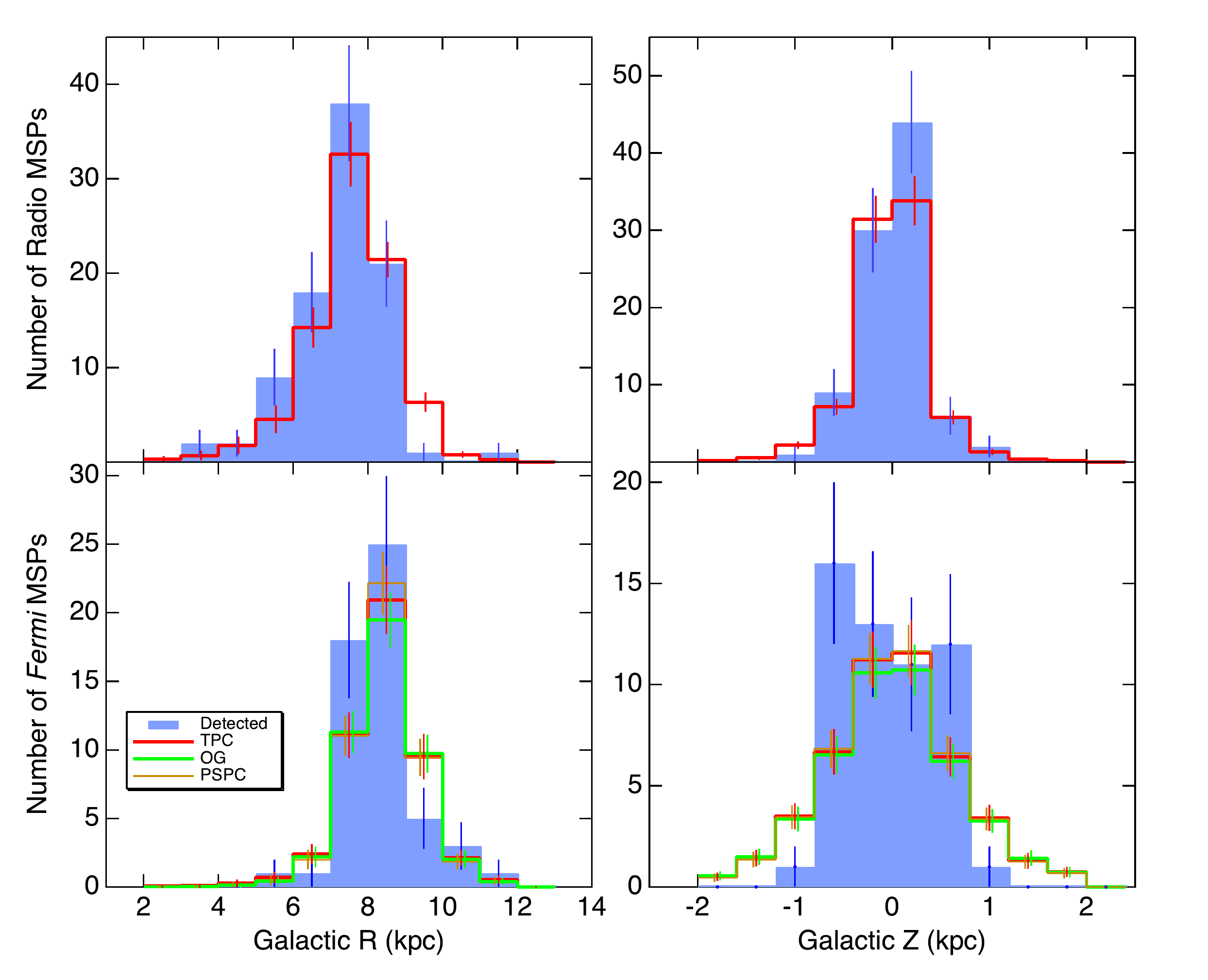}
\caption{Distributions of the Galactic radius $R$ (left) and distance $Z$ out of the Galactic disk (right) of detected (solid blue histograms) of radio MSPs (top panels) and \fermi MSPs (bottom panels) with Poisson uncertainties.   In the top panels simulated radio MSPs are indicated in open red histograms.  In the bottom panels simulated \fermi MSPs are shown using the high-energy emission geometries TPC (open red histograms), OG (open green histograms) and PSPC (open brown histograms). The simulated histograms are obtained by averaging over all the accepted steps after a burn-in period of the MCMC simulation, while their uncertainties represent the standard deviations.}
\label{fig:GalacticRZ}
\end{center}
\end{figure}

In Figure \ref{fig:Aitoff}, we present Hammer-Aitoff projections for the detected and simulated radio and \fermi MSPs.  In general, the MSPs tend to be closer to us than normal pulsars, and display a much larger out-of-plane distribution than normal pulsars.  While the number of pulsars is limited, the simulated distributions do seem to agree with those detected.  
In Figure \ref{fig:GalacticRZ}, the distributions of radio and \fermi MSPs  in the Galaxy are presented as histograms of the Galactic radius $R$ and the distance from the Galactic disk $Z$ both in kiloparsecs with the uncertainties predicted by the high-energy models slighted displaced from each other.  While the distance from us of the simulated MSPs is known with precision, the distance of the detected MSPs are not well known, as the distances are obtained from the detected dispersion measure, assuming an electron density model \citep{CorLaz03}.  Despite the uncertainties, the Galactic distributions of simulated radio  and \fermi MSPs seem to agree  well with those detected.
\begin{figure}[h!]
\begin{center}
\includegraphics[trim=0.5in 0.in 0in 0in, scale=0.45]{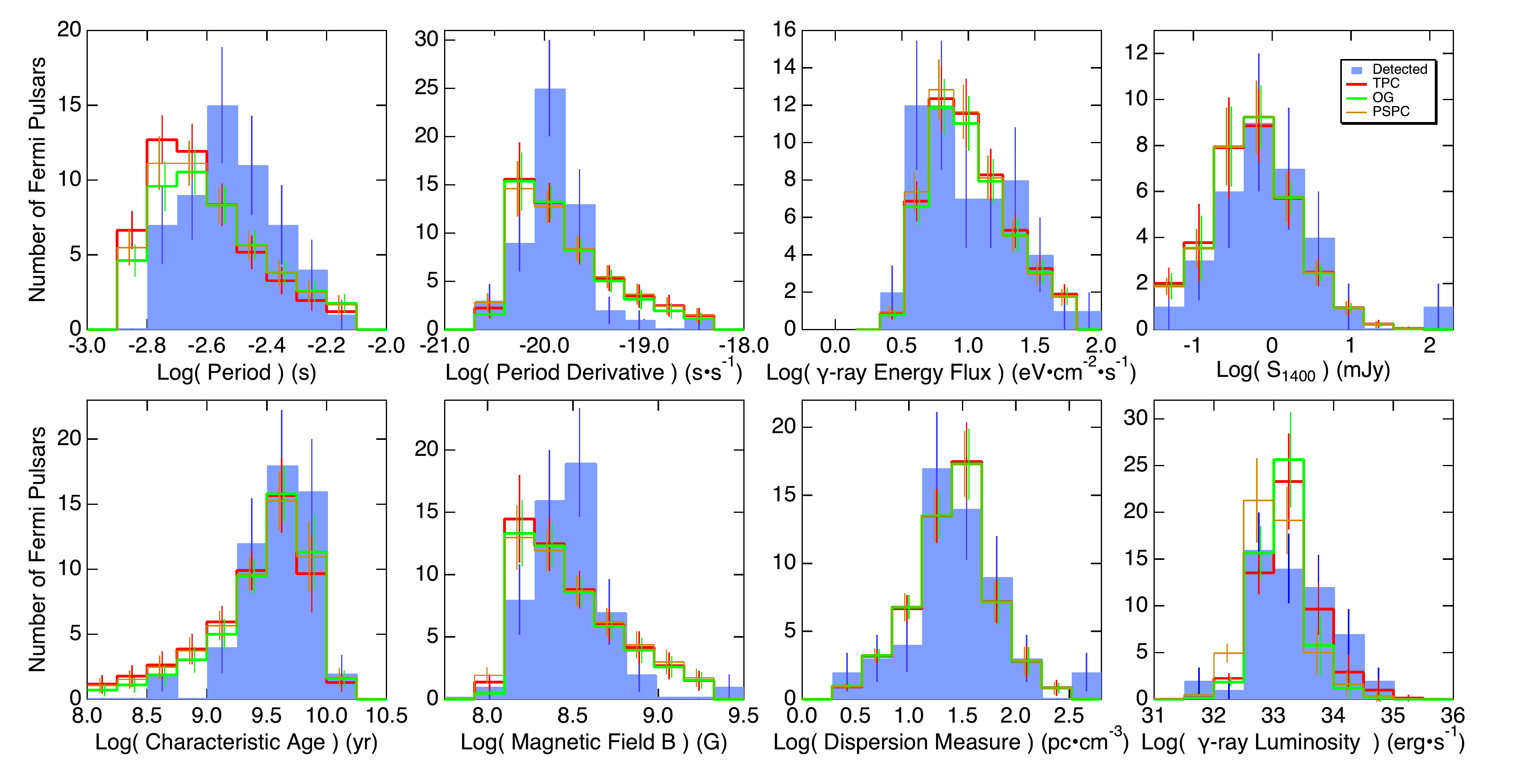}
\caption{Distributions of indicated \fermi pulsar characteristics for detected (solid blue histograms)  with Poisson uncertainties and for simulated using the high-energy emission geometries TPC (open red histograms), OG (open green histograms) and PSPC (open brown histograms). The simulated histograms are obtained by averaging over all the accepted steps after a burn-in period of the MCMC simulation, while their uncertainties represent the standard deviations.}  
\label{fig:Fermi}
\end{center}
\end{figure}

A set of indicated characteristics of \fermi MSPs are compared with those simulated in Figure \ref{fig:Fermi} for the three high-energy emission models.  The nearly identical results predicted by the three high-energy models suggests that there is not clear signature to distinguish among these high-energy emission models.  Generally the agreement by eye is fair with the exception of the magnetic field distribution.  The simulation predicts too many short period \fermi MSPs; however, in the case of the radio MSPs in Figure (\ref{fig:Radio}), the simulated period distribution agrees well with the one detected.  

In the subsequent simulated scatter plots being compared to those of detected MSPs, we select the most likely values of the parameters of the luminosity models indicated in Table \ref{tab:paraunc} for the TPC high-energy model.  Since the MSP birth rate is fixed, the simulation predicts numbers of radio and \fermi MSPs that are slightly different than the detected 92 radio MSPs and 54 \fermi MSPs in the 1FGL.

\begin{figure}[h!]
\begin{center}
\includegraphics[trim=0.5in 0.5in 0in 0in, width=5.in]{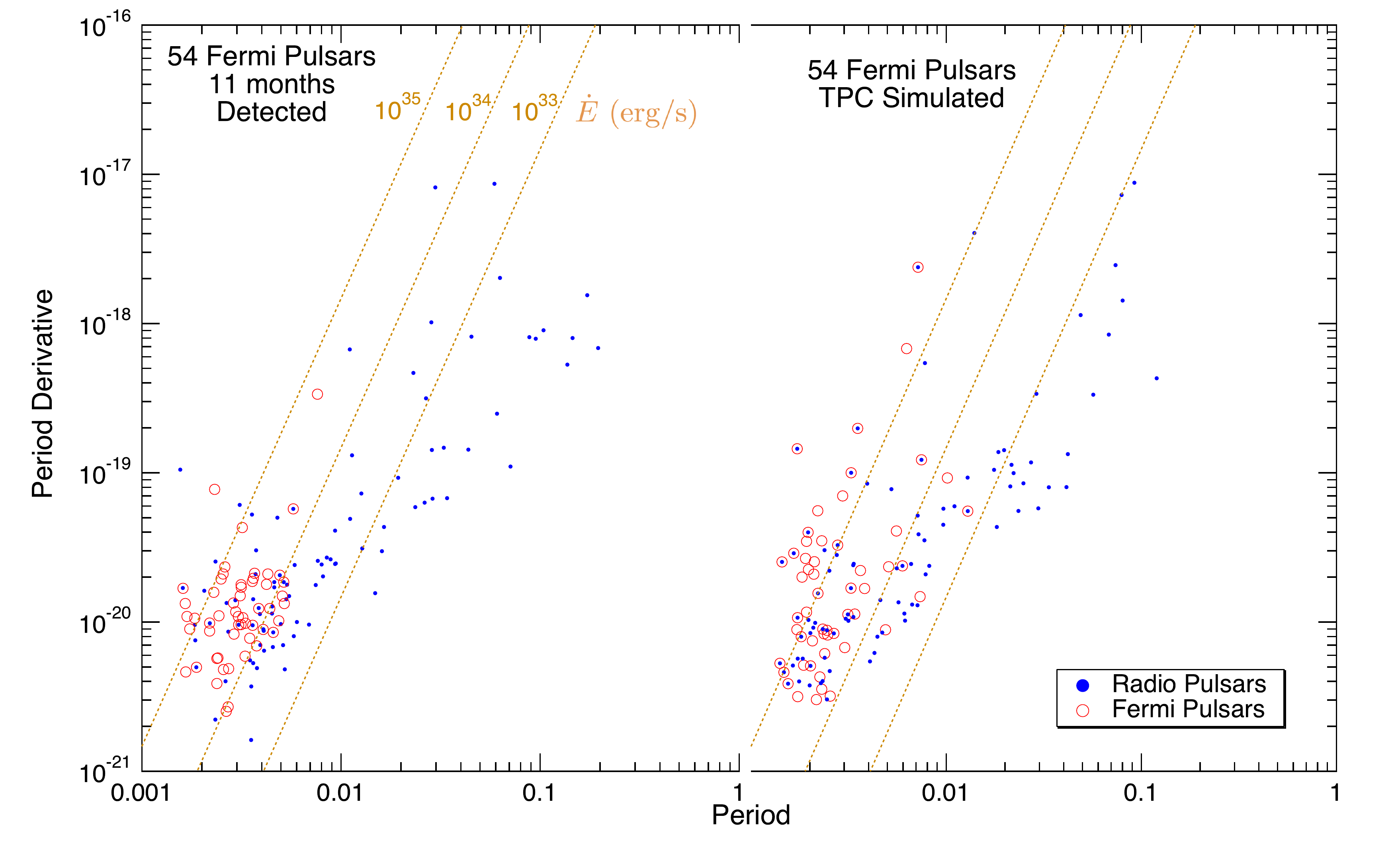}
\caption{Distribution of detected (left panel) and simulated (right panel) of radio (solid blue circles) and \fermi (open red circles) MSPs in the $\dot P-P$ diagram assuming the TPC high-energy emission model. The simulation was performed using the values of the parameters of the luminosity models in Table \ref{tab:paraunc} that maximize the corresponding likelihoods.}
\label{fig:PdotP}
\end{center}
\end{figure}

The distributions of MSPs in the period derivative - period $\dot P - P$ plot are presented in Figure \ref{fig:PdotP} for the detected (left panel) and simulated (right panel) radio (blue solid dots) and \fermi (open red circles).  The period derivatives of the simulated pulsars have the added small contribution from the proper motion (Shklovskii effect \citep{Shklo70}) so that we can compare the simulated $\dot P$ values to those of the \fermi MSPs that have not been corrected for the Shkloskii effect.   For this figure and the next ones, we have used the values of the parameters of the luminosity models that maximize the likelihood indicated in Table \ref{tab:paraunc}.  Note that the 92 detected radio pulsars selected for the study are those that were seen in a group of 13 radio surveys with only 11 of these radio pulsars seen as point sources by \fermi in the 1FGL.  With the exception of these 11 MSPs and one detected through a \fermi blind search, the other \fermi MSPs were discovered through dedicated pointed radio observations of \fermi MSP candidates.  Since the predictions of the assumed high-energy emission models result in nearly identical distributions of the properties of MSPs as seen in Figure (\ref{fig:Fermi}), as an example, we show the simulated distribution in the $\dot P - P$ diagram from the TPC high-energy emission model.  The overall shape of the distributions of the detected MSPs differ somewhat from those simulated with the distribution of the detected \fermi MSPs being more confined in $\dot P$ than the simulated distribution, and the distribution of the detected radio MSPs being more dispersed in $\dot P$.  The power law model of Equation (\ref{eq:Bf}) mainly governs the distribution of MSPs in the $\dot P- P$ diagram along with the various selection effects.    While the agreement of the simulated distributions of the radio MSPs in the $\dot P-P$ diagram in Figure \ref{fig:PdotP} and in Figure \ref{fig:Radio} are acceptable, those predicted for the \fermi MSPs have significant   differences.  The simulation over predicts the short period \fermi MSPs with small $\dot P$ values.  The parameters in Equation (\ref{eq:Bf}), $B_{\rm min}, B_{\rm max}$, and the index $\alpha_B$ where not searched for in the MCMC simulation.  Perhaps a compromise of these parameters might lead to improved agreement of the \fermi distributions.  Performing a sensitivity analysis, we find that these simulated histograms are not significantly affected by the assumed birth distributions in Section 3.  We varied by 20 \% the assumed values of the parameters describing the birth distributions and find there is no significant difference in the simulated histograms, suggesting that the present-day spatial distribution of MSPs are equilibrated, with the age of the MSPs no longer correlated with its location within the Galaxy.
\begin{figure}[h!]
\begin{center}
\includegraphics[trim=0.0in 0.in 0in 0in, width=5.in]{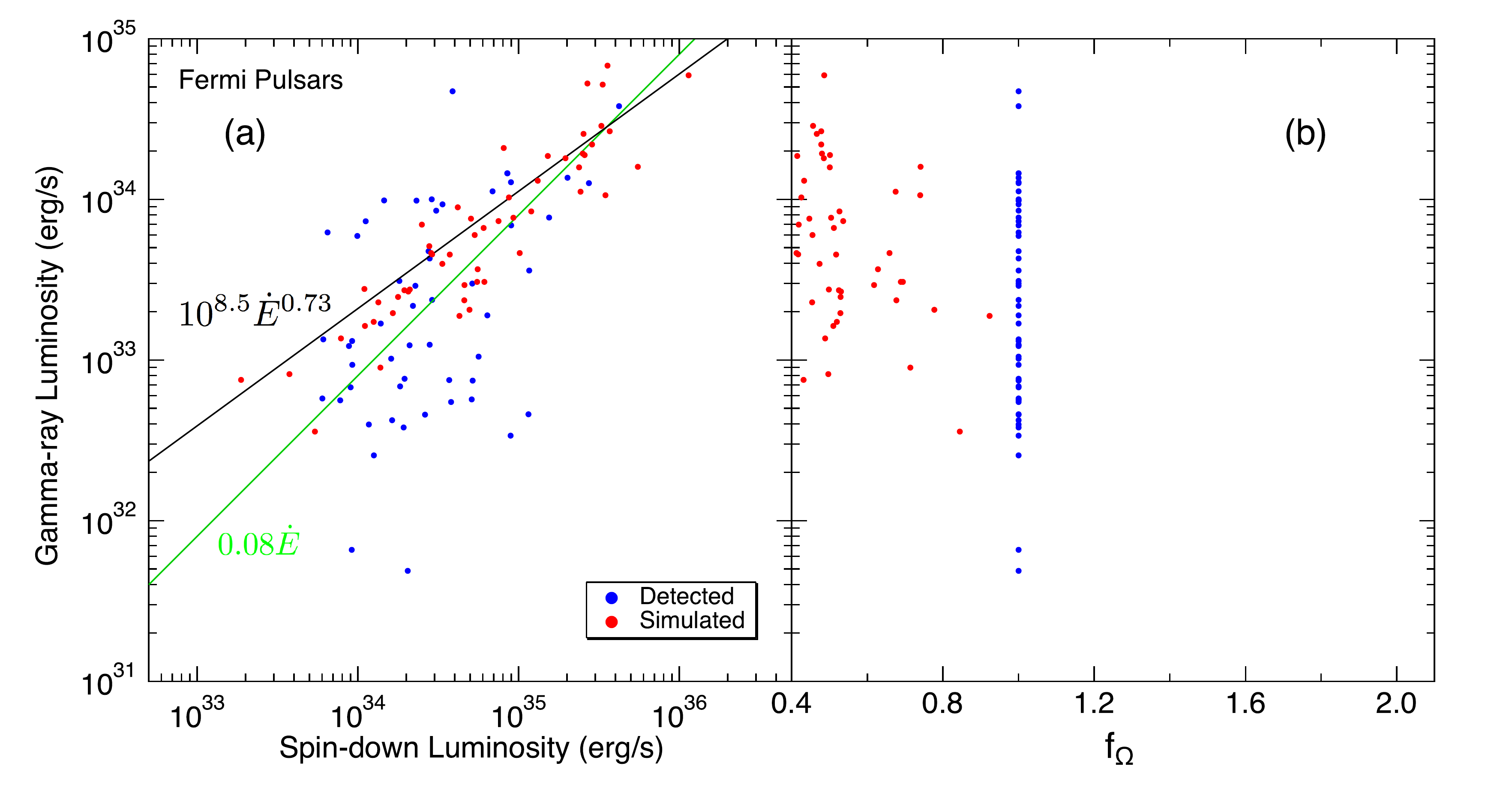}
\caption{The $\gamma$-ray luminosity is plotted versus the spin-down luminosity $\dot E$ (a) and the beam correction factor $f_\Omega$ (b) for the detected (solid blue dots) and the simulated (solid red dots) \fermi MSPs with the TPC high-energy emission model. The simulated $\gamma$-ray luminosities have been rescaled to an $f_\Omega=1$ as assumed for the \fermi MSPs.  The green line in (a) represents an 8\% $\gamma$-ray efficiency and the black line is the TPC high-energy emission luminosity model suggested by the MCMC simulation obtained by a power-law fit to the simulated $\gamma$-ray luminosities and spin-down energies.}
\label{fig:GamLum}
\end{center}
\end{figure}

The governing assumption we are invoking in this study is that MSPs are standard candles.  We assume that the radio $L_\nu$ and $\gamma$-ray $L_\gamma$ luminosities  are power-law functions of the period and period derivative with $\alpha$ and $\beta$ exponents, respectively.  The suggested trend in the \fermi 2PC is that the $\gamma$-ray luminosity of MSPs is more or less proportional to the spin-down energy $\dot E$ (see Figure 9 in \citet{Abdo2PC}), which would suggest $\alpha_\gamma=-3$ and $\beta_\gamma=1$, values different than those in Table \ref{tab:paraunc} obtained with the MCMC simulations.  However, in the 2PC the assumption is that $f_\Omega=1$ and most of the distances determined from the NE2001 electron density using the pulsar's DM.  Part of the spread can be due to different viewing geometries - inclination angles $\alpha$ and viewing angles $\zeta$ resulting in different values of the beaming correction factor $f_\Omega$.  Yet, there are factors of several orders of magnitude spread in the inferred luminosities of the detected MSPs suggesting that other factors might be involved.   In Figure (\ref{fig:GamLum}), we show the plot of the inferred luminosity of the detected MSPs and the model luminosity of simulated MSPs as a function of spin-down energy $\dot E$ for the TPC high-energy emission model.  Since the $\gamma$-ray luminosities of the detected MSPs have been inferred assuming that $f_\Omega=1$, the simulated luminosity has been rescaled to an $f_\Omega=1$ in order to more adequately compare to the \fermi MSPs.  The $\dot E$ of the detected MSPs have been recalculated using the assumed moment of inertia in our spin-down model.    The observed spread in the $L_\gamma$ vs. $\dot E$ of the simulated \fermi MSPs is significant, but not quite as large as for observed in the detected MSPs, perhaps suggesting a deficiency of the TPC $\gamma$-ray emission model.  The black solid line is a result of a power-law fit to the simulated luminosities and spin-down energies $\dot E$, while the green line represents the commonly assumed 8\% $\gamma$-ray efficiency.  The simulated beam correction factors $f_\Omega$ are distributed within a region ranging from 0.4 to 1.  The values of $f_\Omega$, obtained from fitting the radio and $\gamma$-ray light curves of \fermi MSPs in the studies of \cite{Johnson14} and \cite{Venter12} tend to be less than 1 further reducing the $\gamma$-ray luminosities $L_\gamma$.  However, these high-energy emission models do not seem to provide variations of $f_\Omega$ that introduce several orders of magnitude into the $\gamma$-ray luminosities.

So far there has been no real distinction among the assumed high-energy emission models from the MCMC and MC results; the detection sensitivity depends on the simulated phase-averaged flux, ignoring the structure of the light curves.  As mentioned previously, the simulation assumes that the radio light curve has an effective pulse width smaller than 60\% of the phase.
\begin{figure}[h!]
\begin{center}
\includegraphics[trim=0.2in 0in 0in 0in, width=7.5in]{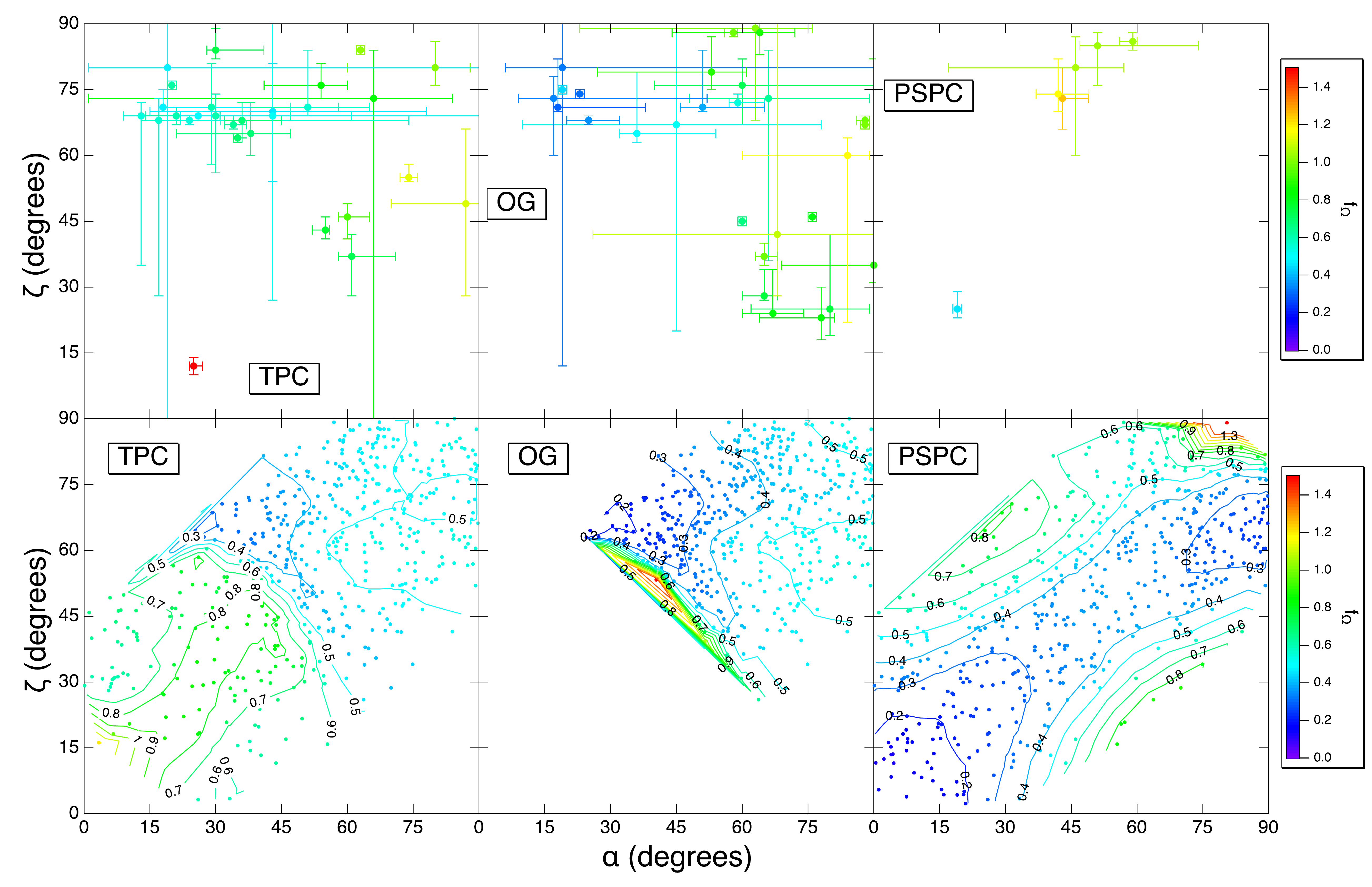}
\caption{The viewing angle $\zeta$ is plotted versus the inclination angle $\alpha$ from the fits of the radio and $\gamma$-ray light curves from Tables 6, 7 and 11 of \citet{Johnson14} (top panels) and of the 540 simulated \fermi MSPs (bottom panels).  The color coding and contour colors represent flux correction factor $f_\Omega$ by the color scale on the right side of the figure. }
\label{fig:ZetaAlpha}
\end{center}
\end{figure}

In Figure (\ref{fig:ZetaAlpha}), the viewing angle $\zeta$ is plotted as a function of the inclination angle $\alpha$ obtained from the Tables 6, 7, and 11 of the work of \citet{Johnson14} (top panels) who fit both the radio and $\gamma$-ray light curves of the \fermi MSPs in the 2PC.  They obtained the best fits for groups of MSPs within each of the high-energy emission models as indicated.   We simulated ten times the number of MSPs required by the assumed birthrate in our groups of radio MSPs (92) and  of \fermi MSPs (54) with the results shown in the bottom panels.  The color coding represents the range in values of the $\gamma$-ray flux correction factor $f_\Omega$ indicated by the color scale on the right side of the panels.  The fits in the top panels indicate a paucity along the line $\zeta=\alpha$.  In contrast, the simulation does not indicate any difficulty populating this region.  On the other hand, some of the preferred locations of the fits with large $\zeta$ and small $\alpha$ are regions that are not populated by the simulation.  Nevertheless, the error bars resulting from the fits are quite large.  The results of the fits tend to be in the upper right half of the $\zeta$ vs. $\alpha$ region that is equally populated by the three high-energy emission models, suggesting that they are equally probable and there is no preferred model.  However, there are noticeable differences in the $\zeta$ vs. $\alpha$ space populated by $\gamma$-ray pulsars for the various assumed high-energy emission models.  There is a large region for $\alpha < 50^\circ $ that is excluded for the OG model as seen previously in the purple region in Figure (\ref{fig:Foms}). 

 However, a region with the smallest $f_\Omega$ in the TPC and OG models is located in the upper left corner of large $\zeta$ and small $\alpha$ in the upper panels.  Yet, the simulation does not populate this region due to the imposed requirement that the \fermi MSPs have a radio flux above $30\ \mu$Jy and a dispersion measure larger than 2.5 $\rm pc/cm^3$ due to the required radio pointed observations that discovered their pulsations.  In performing the light curve fits, \cite{Johnson14}  considered the intensity of the light curves in arbitrary units and did not take into account the associated radio and $\gamma$-ray fluxes to obtain absolute intensities, and so did not have the radio flux constraint we have imposed in this study.  The Figure (\ref{fig:ZetaAlpha})  suggests that some of these fits would result in very weak radio fluxes that perhaps would not be detected, suggesting that light curving fitting procedures need to also constrain the parameter space with the absolute intensities of the radio and $\gamma$-ray light curves of the pulsars being fit.

\subsection{Radio-weak \fermi MSPs}
As discussed previously, we required that our simulated pulsars have radio fluxes above $30\ \mu$Jy and a dispersion measure (DM) larger than 2.5 $\rm pc/cm^3$ in order for them to be realistically discovered as MSPs through dedicated radio observations.  Our model parameters were fixed by matching characteristics of 92 radio MSPs detected in a group of thirteen surveys and of 54 \fermi MSPs that were originally seen as point sources in the 1FGL for an observing period of 11 months.  However, relaxing the radio flux $S_{1400}$ and DM restrictions, we find that the simulation predicts 49 \fermi radio-weak MSPs that are below $30\ \mu$Jy, which is approximately equal to the predicted 52 \fermi radio-loud MSPs.  
\begin{figure}[h!]
\begin{center}
\includegraphics[trim=0.in 0in 0in 0in, width=4.in]{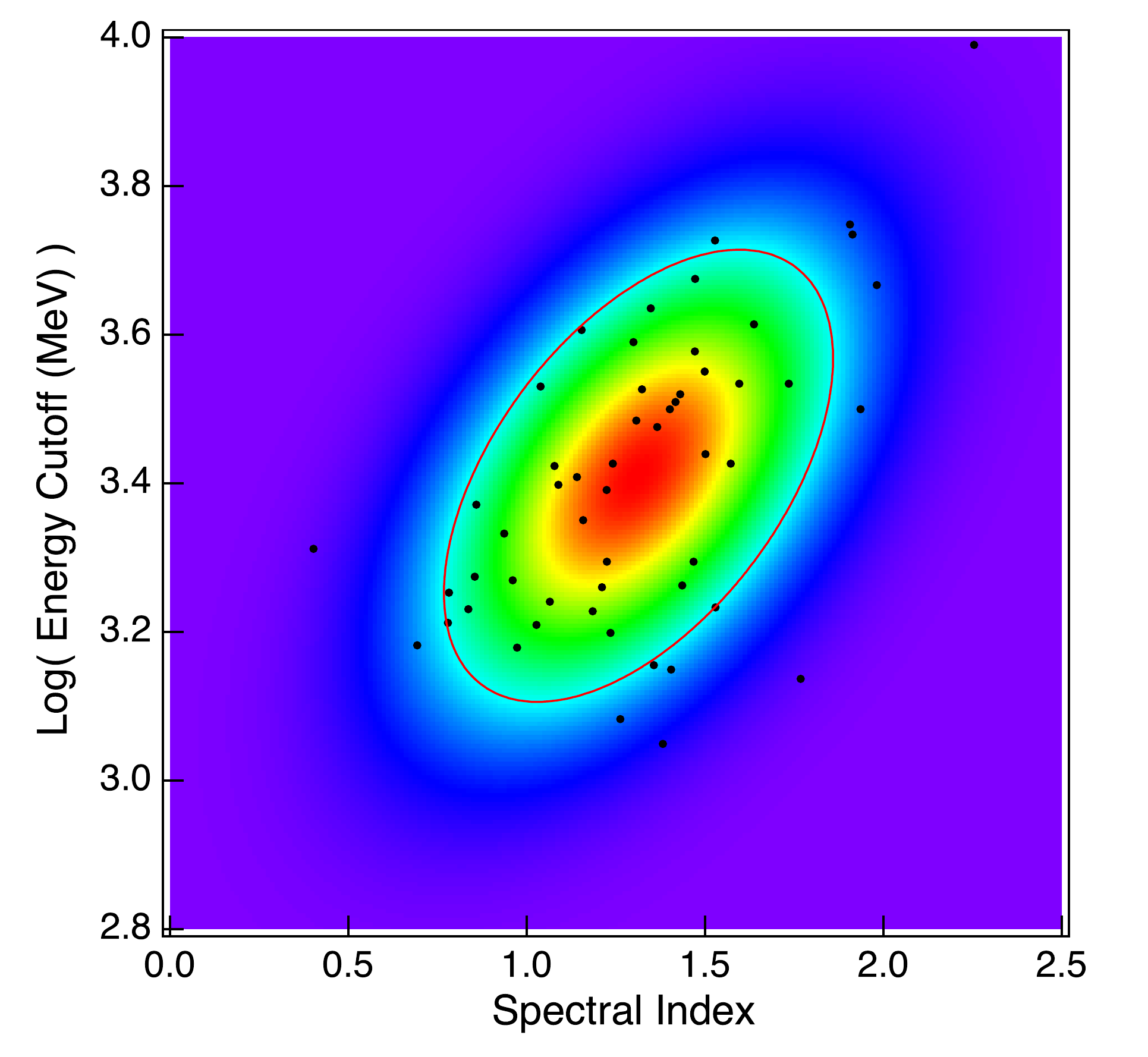}
\caption{The logarithm of the energy cutoffs is plotted versus the spectral indices of the 54 \fermi MSPs shown as black solid circles.  A 2-D Gaussian distribution, shown as the image, is used in the simulation code to randomly generate the energy cutoffs and spectral indices of simulated MSPs.  The red contour represents the 1 $\sigma$ probability contour. }
\label{fig:EcutSpec}
\end{center}
\end{figure}

\begin{figure}[h!]
\begin{center}
\includegraphics[trim=0.in 0in 0in 0in, width=7.in]{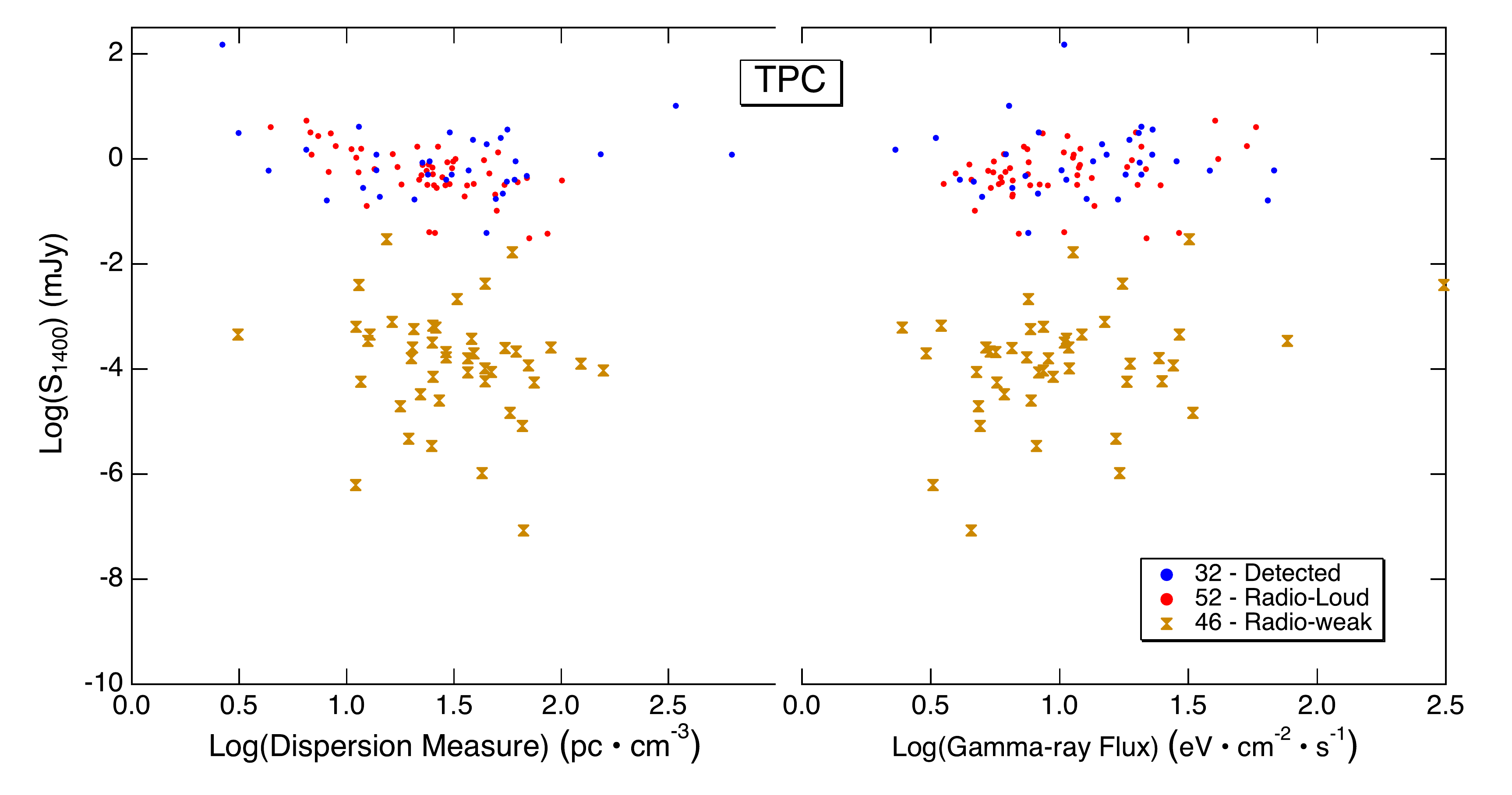}
\caption{The logarithm of the radio flux $S_{1400}$ is plotted versus the logarithm of the dispersion measure and the logarithm of the $\gamma$-ray flux.  The  \fermi detected MSPs that have reported $S_{1400}$ are represented with 32 solid blue dots.  52 simulated radio-loud MSPs with radio fluxes above $30\ \mu$Jy are shown as solid red dots while 46 simulated radio-weak with radio fluxes below $30\ \mu$Jy are shown as solid brown hourglasses.  The simulation was performed with the TPC high-energy emission model.}
\label{fig:S1400DM}
\end{center}
\end{figure}

In Figure (\ref{fig:S1400DM}) the $\log\left(S_{1400}\right)$ is plotted as a function of the $\log\left(DM\right)$ and of the logarithm of the $\gamma$-ray flux for the detected (solid blue dots), simulated radio-loud (radio fluxes above $30\ \mu$Jy) (solid red dots), and simulated radio-weak (radio fluxes below $30\ \mu$Jy) \fermi MSPs (brown hour glasses).  Assuming the TPC high-energy emission model, the simulation predicts an additional 49 MSPs that \fermi sees as point sources with radio fluxes at 1400 MHz below $30\ \mu$Jy shown as solid brown hour glasses. Yet, the MSPs in this group of simulated radio-weak MSPs have comparable $\gamma$-ray fluxes to those of the radio-loud group.   Eleven of these radio-weak MSPs have radio fluxes about $1\ \mu$Jy while 33 radio-weak MSPs have radio fluxes above $0.1\ \mu$Jy.  For a given radio luminosity and distance, larger radio fluxes occur for $\zeta\,=\,\alpha$ or when the impact angle $\beta=\zeta-\alpha$ is small.  However, this is not necessary true for $\gamma$-ray fluxes in the TPC high-energy emission model as seen in Figure 2 where there is a significant green region with $f_\Omega=1$ away from $\zeta\,=\,\alpha$.  It is these MSPs with large $\beta$ where the $\gamma$-ray flux can be significant and the radio flux significantly reduced as the line of site $\zeta$ crosses the outer edge of the cone beam.  As the impact angle $\beta$ increases, the contribution of the cone beam decreases when $\beta$ is larger than the annulus of the cone beam.  For many of these short period MSPs, the core beam is broader than the annulus of the cone beam, and the core beam begins to contribute more strongly than the cone beam,  accounting for the group of weak fluxes $S_{1400}$ around $10^{-4}$ mJy seen in Figure (\ref{fig:S1400DM}).  Many of these \fermi radio-weak MSPs have fluxes below the sensitivity of the deep searches that have been conducted by the \fermi Pulsar Search Consortium.  Their discovery will have to await deeper radio searches of \fermi MSP candidates.

\subsection{Projections of \fermi-LAT MSP detections within other periods of observation}  \label{sec:Proj}
Having fixed the parameters of the luminosity models through MCMC simulations using the radio pulsars detected in thirteen surveys and the \fermi pulsars detected as point sources in the 1FGL, we have made projections for the various observing periods indicated in Table \ref{tab:Proj}.  We have run MC simulations using the accepted MCMC steps, adjusting the \fermi point source threshold map for each of the viewing periods.  The estimated numbers and their uncertainties indicated in Table \ref{tab:Proj} are the averages and standard deviations of the predicted numbers by the MCMC accepted steps for each high-energy emission model.    We indicate the recently updated numbers of detected \fermi radio-loud and radio-weak MSPs for the 2FGL \citep{Nolan12} and 3FGL \citep{Acero15} catalogs, including some radio-weak \fermi MSPs recently discovered using the FAST observatory \citep{Nan2006}, the Einstein@Home program \citep{Wu2018}, and various deep observations at frequencies lower than 1400 MHz.  In addition, we project the number of MSPs to be detected  for the 5 and 10 year observing period where \fermi-LAT is expected to detect around 120 and 170 MSPs.   The $\gamma$-ray emission of the PSPC model takes place over the open-field volume, and so tends to be more aligned with the radio beams, leading to fewer \fermi radio-weak MSPs.  The numbers for the high-energy emission geometries of the TPC and OG models are very similar with somewhat more restricted emission over the sky in the OG model.  As can be seen in Figure (\ref{fig:ZetaAlpha}) there is a larger viewing region available for the TPC model than the OG model.  The \fermi radio-weak MSPs fill-in the regions that are not populated by the \fermi radio-loud MSPs in this figure, therefore there are significantly more \fermi radio-weak MSPs associated with the OG model.  
\begin{deluxetable}{c|c|c|c|c|c|c|c|c|c}[h!]
\tablecolumns{10}
\tablewidth{0pc}
\tablecaption{Predicted \fermi-LAT future detections \label{tab:Proj} }
\tablehead{
\colhead{Catalog} & \colhead{Observing Period} &  \multicolumn{2}{c}{Detected } & \multicolumn{6}{c}{Simulated }}
\startdata
		&			&	&	& \multicolumn{2}{c}{TPC} &  \multicolumn{2}{c}{OG} &  \multicolumn{2}{c}{PSPC} \\ \hline
		&			& radio-loud &	radio-weak	&	radio-loud	&	radio-weak	&radio-loud	&	radio-weak	&radio-loud	&	radio-weak	\\ \hline
BSL		& 3 months	& 12		   & 	1		& $26\pm 3$ 	&	$21\pm 3$	& $25\pm 3$	 &	$32\pm 4$	& $26\pm 3$ 	&	$7\pm 1$ \\
1FGL 	& 11 months 	& 54 		   & 	7		& $52\pm 5$ 	&	$46\pm 6$ 	& $50\pm 5$ 	&	$66\pm 8$	& $53\pm 5$ 	&	$16\pm 2$ \\
2FGL 	& 2 years 		& 64    	   & 	10		& $78\pm 7$ 	&	$71\pm 8$ 	& $75\pm 7$ 	&	$101\pm 12$ 	& $80\pm 7$ 	&	$26\pm 4$\\
3FGL 	& 4 years 		& 87		   & 	11		& $111\pm 10$ 	&	$104\pm 12$ 	& $105\pm 10$ &	$145\pm 14$ 	& $ 114\pm 10$ &	$40\pm 5$ \\
		& 5 years 		&		   & 				& $124\pm 11$ 	&	$117\pm 14$ 	& $116\pm 11$ 	&	$163\pm 19$ 	& $127\pm 12$ &	$45\pm 6$ \\
		& 10 years 	& 		   & 				& $173\pm 15$ &	$171\pm 19$	& $160\pm 14$ &	$232\pm 27$	& $178\pm 16$ &	$69\pm 9$ \\
\enddata 
\end{deluxetable}

\subsection{The contribution of MSPs to the  diffuse $\gamma$-ray background}
We find that the three high-energy emission models, TPC, OG, and PSPC, do not display significantly different signatures of the directly observable properties of MSPs, as previously shown in Figure (\ref{fig:Fermi}).  The light curves from the TPC model are more similar to those from emission geometries of more realistic descriptions of pulsar magnetospheres (\cite{Kalapotharakos18,Brambilla18}, and references therein).  Therefore, in the sections to follow, we focus on the results of the simulations assuming the TPC high-energy emission model.  

We now consider the contribution of MSPs to the Galactic and isotropic diffuse $\gamma$-ray background by keeping track of characteristics of MSPs that are below the \fermi point source threshold, using the TPC high-energy emission model.  However, such comparisons required the conversion of the energy flux into photon flux, entailing the estimation of the spectral indices and energy cutoffs of the simulated MSPs.  In the \fermi 2PC a mild correlation was found between the spectral indices and $\dot E$ and between the cutoff energies $E_{\rm cut}$ and the magnetic field at the light cylinder.  Here we chose to relate the spectral indices and cutoff energies in a similar manner as performed in the study of NPs by \cite{Pierbattista12}.  We used a 2-D correlated Gaussian distribution to represent the observed distribution of the logarithm of the cutoff energies $E_{\rm cut}$ and spectral indices $\alpha$ of MSPs.  In Figure (\ref{fig:EcutSpec}), the cutoff energies and spectral indices of the 54 \fermi MSPs are plotted.  Given the cut-off energies $\log E_{cut}$ and the spectral indices  $\epsilon$ of the MSPs, we generated the the probability distribution using  Kernel Density Estimation \citep{Scott15} with a 2-D Gaussian kernel.  The resulting probability distribution  had  the following set of statistical parameters: a mean $\mu_{\rm cut} = 3.41$, a width $\sigma_{\rm cut} = 0.21$ of the $\log E_{\rm cut}$, a mean $\mu_\epsilon = 1.32$, a width $\sigma_\epsilon=0.37$ of the spectral indices, and a correlation coefficient $\rho=0.52$, which we chose to obtain the 2-D Gaussian shown as an image plot in the Figure (\ref{fig:EcutSpec}).   Assuming a power law with an exponential cutoff, one can obtain the photon flux  of each simulated pulsar from the energy flux, the cutoff energy and the spectral index obtained from random draws from the correlated 2-D Gaussian distribution with the above established parameters.  The same correlated 2-D Gaussian distribution of $\log E_{\rm cut}$ and spectral index $\epsilon$ is used for the Galactic Center Excess and globular clusters in the following sections.
\begin{figure}[h!]
\begin{center}
\includegraphics[trim=0.2in 0in 0in 0in, width=4.in]{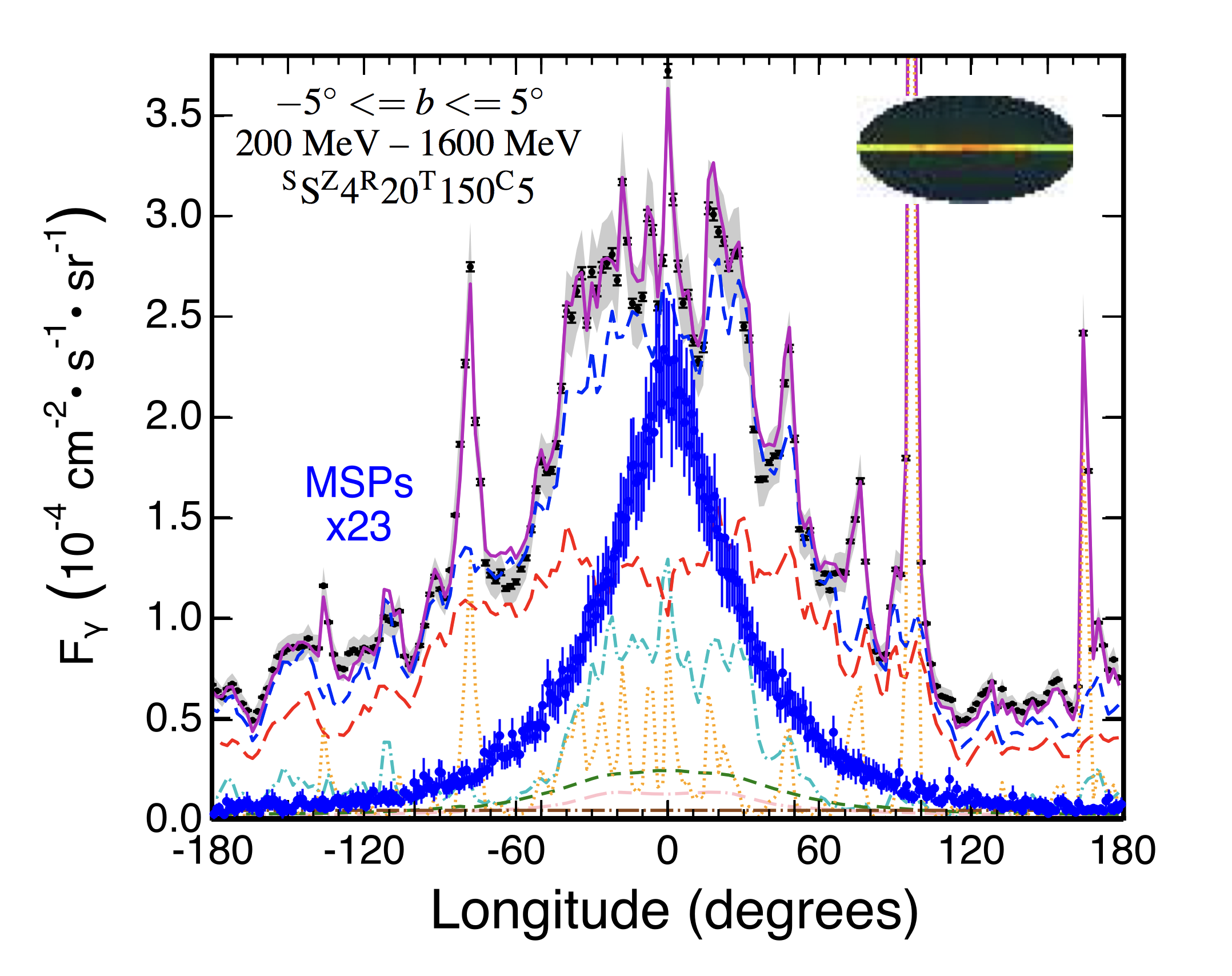}
\caption{ The contribution of MSPs to diffuse $\gamma$-ray background from the Galactic Plane (from the upper left panel of Figure 21 of \cite{Ackermann2012}).  The flux of the simulated MSPs has been multiplied by a factor of 23 to roughly provide an upper limit of the contribution from MSPs.} 
\label{fig:DiffPlane}
\end{center}
\end{figure}

\begin{figure}[h!]
\begin{center}
\includegraphics[trim=0.2in 0in 0in 0in, width=5.in]{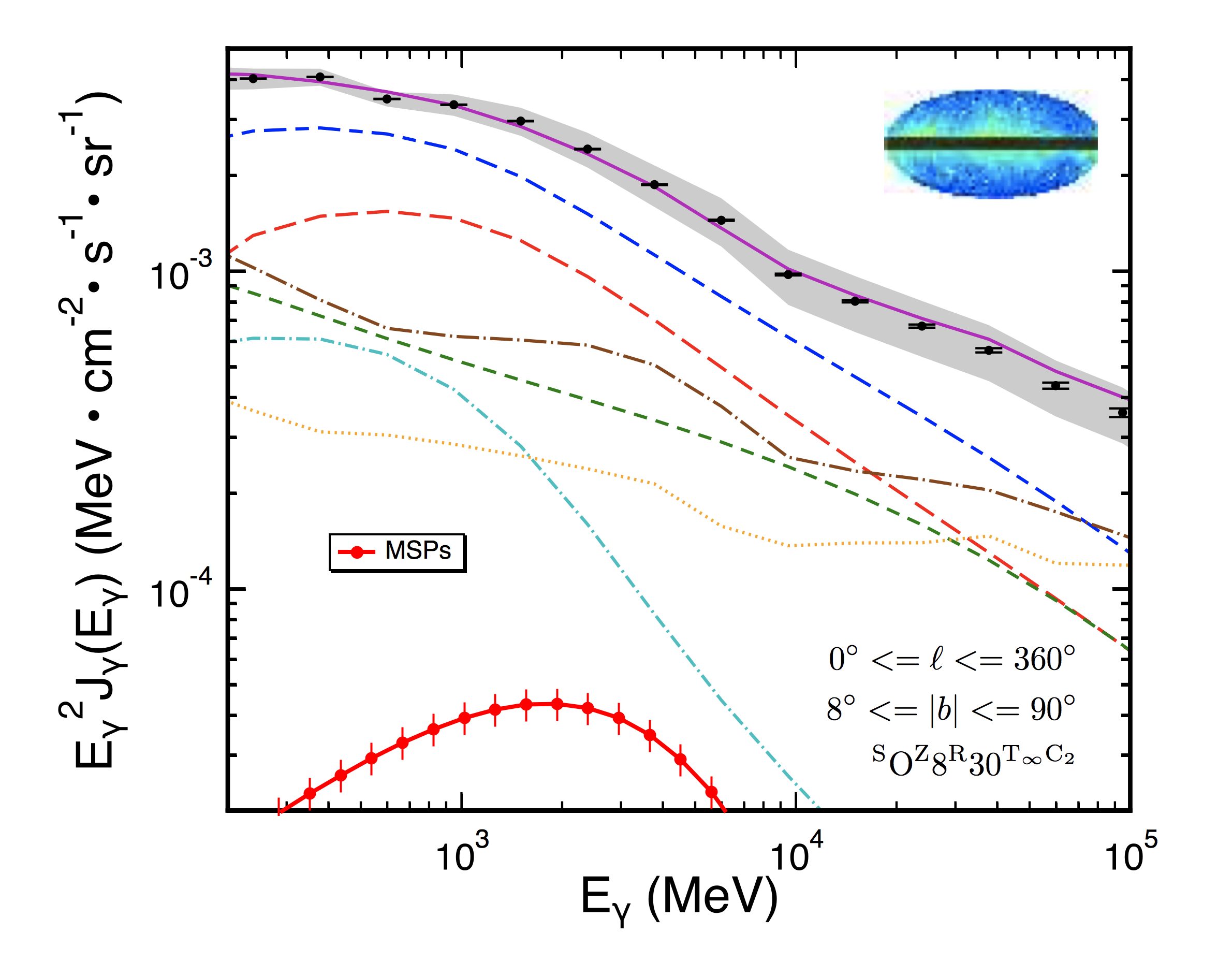}
\caption{The contribution of MSPs (solid red curve) to \fermi diffuse $\gamma$-ray background from the Galactic out-of-plane ($8^\circ\,\leq \, b \, \leq\, 90^\circ$) \citep{Ackermann2012}.  The flux of the simulated MSPs (thick solid red curve) contributes less than $\sim 2\%$ to the out-of-plane diffuse $\gamma$-ray emission. }
\label{fig:OutPlane}
\end{center}
\end{figure}

In Figure (\ref{fig:DiffPlane}), the contribution of the MSPs to the diffuse $\gamma$-ray background from the Galactic Plane has been obtained with the same window in Galactic latitude ($-5^\circ\,\leq\, b\, \leq 5^\circ$) and in the same $\gamma$-ray energy range (200 MeV --- 1600 MeV) as in  \cite{Ackermann2012}.  Assuming the viewing period of 21 months and the TPC high-energy emission model, the simulated flux from MSPs has been multiplied by a factor of 23 to compare its shape to the angular distribution shown in the upper left panel of Figure 21 from \cite{Ackermann2012}, normalizing the contribution from MSPs to the smooth contribution of $H_{\rm II}$ (blue, long-dash-dashed curve) to provide a crude estimate of the upper estimate of the contribution from MSPs.  The uncertainties of the simulated angular distribution result from running the MC code with the accepted MCMC steps as discussed previously. The factor of 23 suggests that the simulated MSPs from the Galactic Disk roughtly contribute less than 4\% to the diffuse Galactic $\gamma$-ray background, which is significantly larger than that estimated in the study of \cite{Calore14} of 0.9\% at 2 GeV.  In addition, we compare in Figure (\ref{fig:OutPlane}) the out-of-plane $\gamma$-ray spectrum of simulated MSPs to the Figure 12 in \cite{Ackermann2012}.  The contribution is less than 2\% to the overall spectrum, which is roughly in agreement with the in-plane contribution, but significantly smaller than the conclusions of \cite{Faucher10} of 5 --- 15\% contribution from MSPs to the high-latitude $\gamma$-ray background.   Given the large width of the \fermi MSP distribution in Galactic latitude, one might expect a larger contribution to the out-of-plane diffuse $\gamma$-ray spectrum from MSPs than to that of the in-plane spectrum; however, there are more undetected $\gamma$-ray MSPs in-plane of the GD than out-of-plane as suggested by the $Z$ distribution of radio MSPs in Figure (\ref{fig:GalacticRZ}).  The $Z$ of radio MSPs is narrower than that of the \fermi MSPs, suggesting that the $Z$ distribution of radio MSPs might be more representative of the distribution of undetected $\gamma$-ray MSPs. The larger number of undetected  $\gamma$-ray, as well as radio, MSPs in-plane of the GD is not entirely unexpected given the larger background interferences in both radio and $\gamma$-ray wavelengths, making it more difficult for \fermi to resolve the source as a point source and for radio observations with the increase in sky temperature, increasing the $S_{\rm min}$ flux threshold of the radio telescope.

 \subsection{\fermi millisecond pulsars in globular clusters}
 Having simulated MSPs from the Galactic Disk, we test our model with a group of globular clusters.  We assume that the MSPs from globular clusters, though perhaps having a different evolutionary history, have the same characteristics as those from the Galactic Disk.  We simulate a large number of MSPs within the cluster to obtain a smooth spectral energy distribution (SED) of all of those MSPs within the cluster.  None of the individual MSPs are seen as \fermi point sources, although \fermi has detected the emission from a number of clusters as a whole.  By normalizing the simulated $\gamma$-ray SEDs of the globular clusters to the ones detected, we can estimate the number of MSPs within each of the globular clusters.  We compare our model predictions with those made by the \fermi team in the work of  \cite{Abdo10}. 
 
The globular cluster 47 Tucanae (Tuc) is one of the closest to us at a distance of 4.0 kpc \citep{McLaughlin06}, having a similar angular diameter \citep{DaCosta82,Lindsay67} as the full Moon in the sky (a field-of-view (FOV) of about 30 arcminutes) and is known to contain 25\footnote{\url{http://www3.mpifr-bonn.mpg.de/staff/pfreire/47Tuc/\#News}} radio MSPs \citep{Pan16}.  \fermi observed 47 Tuc for a period of 546 days \citep{Abdo10} and obtained a SED from 100 MeV to 10 GeV with a best-fit index of $\Gamma= 1.4\pm 0.3$ and a cutoff energy of $E_{\rm cut}=2.2^{+0.9}_{-0.5}$ GeV.  The measured energy flux was determined to be $f_\gamma=(2.5 \pm 0.3) \times 10^{-11}\ {\rm erg \cdot cm^{-2} \cdot s^{-1}}$ yielding an isotropic luminosity of $L_\gamma=(4.8\pm1.1) \times 10^{34}$ erg/s.  Using the prescription given by the expression
\eqs{\label{eq:etaFm} N_{\rm MSPs} \;=\; \dfrac{L_\gamma}{\langle\dot  E\rangle\, \langle \eta \rangle} }
where $L_\gamma$ is the overall $\gamma$-ray luminosity of the cluster, $\langle \dot  E \rangle=(1.8\pm0.7)\times10^{34}\ {\rm erg\cdot s^{-1}}$ is the adopted average spin-down luminosity and $\langle \eta \rangle=0.08$ is the adopted average $\gamma$-ray efficiency, the \fermi team \citep{Abdo10} obtained $33\pm15$ MSPs in 47 Tuc.

\begin{figure}[h!]
\begin{center}
\includegraphics[trim=0.2in 0in 0in 0in, width=5.in]{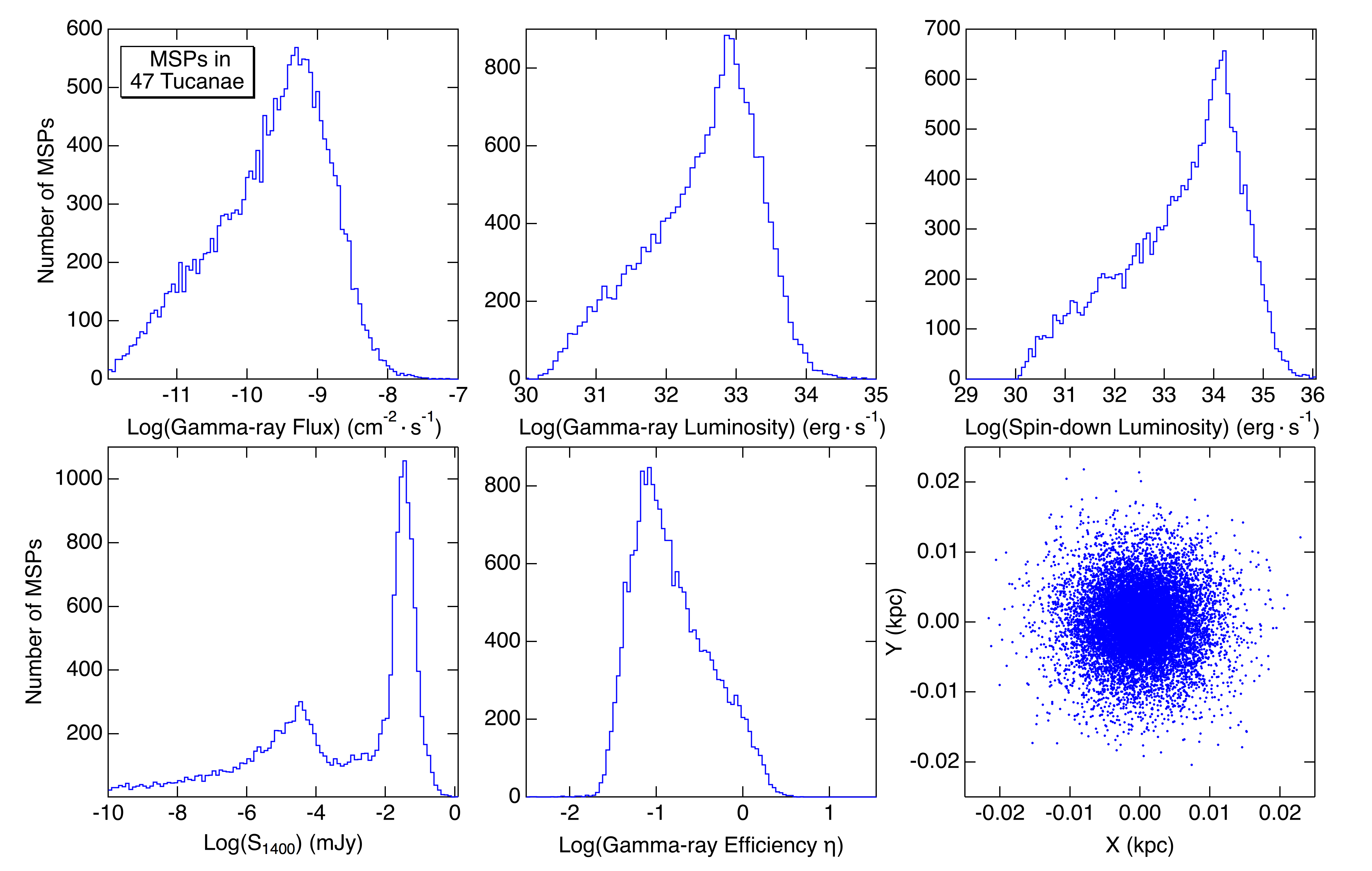}
\caption{Various distributions of indicated characteristics of the simulated MSPs in the globular cluster 47 Tuc.}
\label{fig:47TucHis}
\end{center}
\end{figure}

We performed a simulation of 47 Tuc adapting our code so that the Solar System is 4.0 kpc from a distribution of MSPs to match the distance to 47 Tuc, adjusting the \fermi point source threshold map to the viewing period. We used a density profile of MSPs described by a spherical Gaussian in $r$ with a width $\sigma = 10$ pc to randomly generate MSPs within the cluster to reproduce its angular size mentioned above.  The parameters of the luminosity models were set to the most likely values for the TPC high-energy emission model in Table \ref{tab:paraunc}.
To obtain smooth histograms, we simulated 20,000 MSPs and display in Figure (\ref{fig:47TucHis}) various indicated properties of MSPs.  The histogram of the $\log L_\gamma$ suggests that the MSPs have an average luminosity $\langle L_\gamma \rangle=8.\times 10^{32}$ erg/s, which with the \fermi detected luminosity $L_\gamma\;=\;(4.8\pm1.1) \times 10^{34}$ erg/s, suggests $N_{\rm MSPs} = 60 \pm 14$ MSPs present in 47 Tuc.  The histograms suggest somewhat smaller values of $\langle\dot  E\rangle\;=\; 1.3 \times 10^{34}$ erg/s and $\langle \eta \rangle = 0.08$ than those obtained by the \fermi team.  The only uncertainty included in this estimate is that of the \fermi detected luminosity $L_\gamma$.  Given that the simulation produces MSPs with a wider range of spin-down luminosities than those detected in the GD, our average spin-down luminosity is somewhat smaller than the one estimated in \cite{Abdo10} specific for \fermi MSPs.  The distribution of MSPs in the Y vs. X plot in Figure (\ref{fig:47TucHis}) indicates a diameter ranging from 30 to 40 pc, which at a distance of 4 kpc gives an angular diameter in the range of 25 to 34 arcminutes, in good agreement with its visual size mentioned above.  The results of the simulation are not sensitive to the size of the cluster if the region of interest (ROI) encloses the entire cluster.  

The distributions of the radio flux $S_{1400}$ in Figure (\ref{fig:47TucHis}) display two maxima with the larger one at $S_{1400} \approx 8\ \mu$Jy and the second smaller one at $S_{1400} \approx 0.004\ \mu$Jy.  Since on average there is no variation in distance, the distribution of the radio luminosity (not shown) shows a single peak.  The second peak is caused by large impact angles $\beta\;=\; \zeta - \alpha$ that are greater than the radius of the annulus of the cone beam where the core beam begins to make significant contributions.  The typical radius of the annulus of the cone beam is about $20^\circ$ whereas the width of the core beam is about $29^\circ$.

\begin{figure}[h!]
\begin{center}
\includegraphics[trim=0.2in 0in 0in 0in, width=5.in]{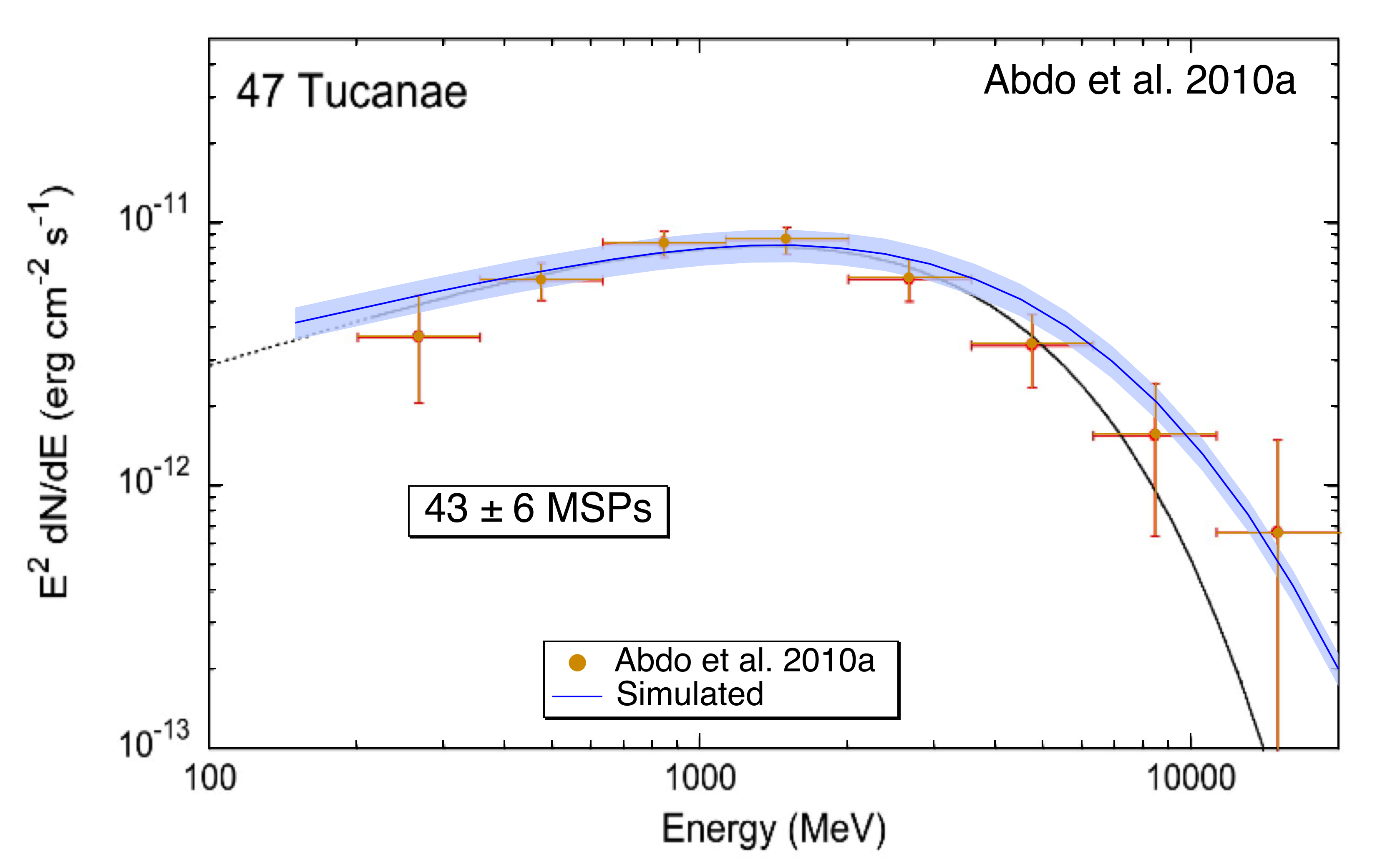}
\caption{The $\gamma$-ray power spectrum of 47 Tuc obtained by the \fermi LAT detector from Figure 2 of  \cite{Abdo10} displayed with a simulated spectrum (blue curve) representing the contribution from $43\pm 6$ MSPs with the shaded light blue region representing 1 $\sigma$ confidence of the simulation.}
\label{fig:47TucSpec}
\end{center}
\end{figure}

\cite{Abdo10} obtained a $\gamma$-ray spectral energy distribution (SED) of 47 Tuc, which we reproduce in Figure (\ref{fig:47TucSpec}) together with a predicted SED from a fit of our simulated SED.    The uncertainties are obtained by running the MC code with the MCMC accepted steps.  With each MC simulation, an average SED is obtained per MSP within the cluster, then for all the accepted MCMC steps, we obtain the average and standard deviation of the SED per MSP. The fit of the simulated SED is performed by adjusting a single multiplicative coefficient, preserving the predicted shape, to the one detected in \cite{Abdo10}.  The simulated spectrum predicts $43{\pm 6}$ MSPs reproduce the observed  SED of 47 Tuc.   This result is in good agreement with our previous number of $60\pm 14$ MSPs together with the estimate by the \fermi team of $33\pm 15$ MSPs \citep{Abdo10}.  It can be seen from the distribution of the radio flux $S_{1400}$, there are a few MSPs predicted to have fluxes above 0.1 mJy and detectable by pointed radio observations.  Normalizing the distribution to 45 MSPs, the simulation predicts 17 and 11 MSPs with radio fluxes $S_{1400} \geq 10\ \mu$Jy and $S_{1400} \geq 30\ \mu$Jy, respectively.  Of the 14 MSPs in 47 Tuc listed in \cite{Camilo00}, all have $S_{1400} \geq 30\ \mu$Jy. 

We simulate the other globular clusters listed in Table 4 of \cite{Abdo10} that have well defined $\gamma$-ray SEDs and summarize our results in Table \ref{tab:Globular}.  The second and third columns labeled $L_\gamma$ and $N_{\rm det}$ contain the cluster luminosities and the number of MSPs estimated by the \fermi team in Table 4 of \cite{Abdo10}, while the fourth column $N_{\rm sim}$ are the predicted number of MSPs simulated for the globular clusters listed by fitting the simulated SED of each of the clusters in a similar manner as described above for 47 Tuc.  
\begin{deluxetable}{lcccc}[h!]
\tablecolumns{4}
\tablewidth{0pc}
\tablecaption{Globular clusters --- \fermi luminosities and expected number of MSPs from Table 4 in \cite{Abdo10} \label{tab:Globular} }
\tablehead{
\colhead{Name} & \colhead{$L_\gamma \left(10^{34}\ {\rm erg\cdot s^{-1}}\right)$ } & \colhead{$N_{\rm det}$ } & $N_{\rm sim}$  }
\startdata 
47 Tucanae  	& $4.8^{+1.1}_{-1.1}$ 	& $33^{+15}_{-15}$ 	& $43\pm 6$  \\
Omega Cen  	& $2.8^{+0.7}_{-0.7}$ 	& $19^{+9}_{-9}$  	& $27\pm 6$ \\
M~62             	& $10.9^{+3.5}_{-2.3}$ 	& $76^{+38}_{-34}$  & $82\pm 15$ \\
NGC~6388   	& $25.8^{+14.0}_{-10.6}$ 	& $180^{+120}_{-100}$  & $220\pm 40$\\
Terzan~5     	& $25.7^{+9.4}_{-8.8}$ 	& $180^{+100}_{-90}$  & $180\pm 30$ \\
NGC~6440  	& $19.0^{+13.1}_{-5.0}$ 	& $130^{+100}_{-60}$   & $130\pm 30$ \\
M~28            	& $6.2^{+2.6}_{-1.8}$ 	& $43^{+24}_{-21}$  & $39 \pm 8$\\
NGC~6652   	& $7.8^{+2.5}_{-2.1}$ 	& $54^{+27}_{-25}$  & $65 \pm 16$\\ \hline
Average		& $0.14^{+0.04}_{-0.03}$ per MSP & $89^{+24}_{-19}$	&	$100 \pm 8$\\
\enddata 
\end{deluxetable}

The estimated number of MSPs listed in the fourth column are a result of normalizing the simulated SEDs to those in \cite{Abdo10} for the other globular clusters as we did for 47 Tuc in Figure (\ref{fig:47TucSpec}).  The predicted numbers of MSPs obtained from the \fermi SEDs for this group of globular clusters are in good agreement with the results of the \fermi team \citep{Abdo10}.  However, our predicted number of MSPs is generally larger than the ones estimated in \cite{Abdo10}, since our total population of simulated MSPs have smaller average luminosity than those of the detected \fermi MSPs.  Roughly about 100 MSPs on average are housed in globular clusters. 
\newpage
\subsection{The Galactic Center Excess}
The Galactic Center (GC) has recently generated great interest as careful studies of the $\gamma$-ray emission suggest an excess (GCE) over the diffuse background from cosmic ray interactions.  Various groups \citep{Gordon13, Hooper13, Calore15a, Calore15b, Abazajian14} and recently the \fermi team \citep{Ajello16} have obtained the spectrum of the GCE, suggesting its origin may come from dark matter annihilation of particles beyond those of the Standard Model.  The GC would then be expected to have the largest density of nearby dark matter.  We focus on the spectrum of the GCE obtained by \cite{Gordon13}(GM13) as it is made available in Appendix A of their study.  They suggest that a mixture of tau particle $\tau^+\,\tau^-$ and b quark $b^+\,b^-$ annihilation is required to adequately fit the \fermi spectrum of the GCE from the decay of a dark matter (DM) particle with a mass between 20 to 60 GeV.  The halo density of DM is expected to follow a Navarro-Frenk-White (NFW) distribution \citep{Navarro96,Klypin02} given by
\eqs{\label{eq:NFWden} \rho(r) \;=\; \dfrac{\rho_{\rm o} }{\left(\dfrac{r}{r_s}\right)^\gamma\left[ 1+ \left(\dfrac{r}{r_s}\right) \right]^{3-\gamma} },}
where $r_s\;=\;23.1$ kpc is the scale radius and $\rho_{\rm o}$ is the scale density that sets the overall DM mass in the halo.
The $\gamma$-ray flux from the annihilation is proportional to the square of the NFW density.  The study of GM13 finds that the exponent $\gamma\;=\;1.2$ provides the best description of the intensity distribution as a function of the distance from the GC.

However, other groups \citep{Wang05, Abazajian11, Yuan14, Macias14} have stressed that the \fermi spectrum of the GCE can be described by the average spectrum of the \fermi MSPs, resulting from a population of MSPs in the GC that are undetected as point sources.   Two groups \citep{Bartels16, Lee16} performed different point source analyses  and concluded that point sources are responsible for the GCE.  The imaging survey of the GC carried out with {\it Chandra} by \cite{Wang02} revealed over 1000 discrete X-ray point sources with many believed to be X-ray binaries. 
\cite{Pfahl02} argued that the survey included about 1\% of the stars within the GC suggesting that many of these sources are wind-accreting neutron stars.  \cite{Muno03} performed a narrower and deeper {\it Chandra} survey of the GC within a region of interest of $17'\times 17'$ around the GC corresponding to an area of $40\times 40$ pc, finding 2357 X-ray point sources, concluding that more than half of these sources have spectral power laws similar to magnetically accreting white dwarfs and wind-accreting neutron stars.  The study of the inner ($r<12'$) bulge of the Andromeda galaxy M31 by \cite{Voss07a} suggested that many of these X-ray point sources are low mass X-ray binaries (LMXBs), which are believed to be the progenitors of MSPs.  \cite{Tremaine75} (revisited by \cite{Gnedin14}) investigated the formation of the nuclear star cluster in M31 with a dynamical friction model involving the tidal disrupting of globular clusters that pass near the center. \cite{Ackermann17} found a $\gamma$-ray excess in the M31 bulge and suggested that it may be due to a population of MSPs.  \cite{Gnedin14} as well as more recently \cite{Brandt15} have suggested that the GCE is a result of MSPs in the GC that have come from disrupted globular clusters.

We explored the possibility of an unresolved population of MSPs within the GC with two possible MSP probability distributions, one given by an approximation of the above Eq. (\ref{eq:NFWden}) with the expression
\eqs{\label{eq:NFW}
P_{\rm NFW}(r)\,dr \;=\; 0.6\, \left( \dfrac{r^{-2.4}}{r_{\rm max}^{0.6}}\right) \, r^2\,dr ,}
where $r$ is in spherical coordinates and $r_{\rm max} = 2.0$ kpc, and a second Gaussian form by
\eqs{\label{eq:Gau}
P_{\rm Gauss}(r)\, dr \;=\; \dfrac{2}{\sigma_R^2}\,e^{-r^2/\sigma_R^2}\, r\, dr}
where $r$ is in cylindrical coordinates and $\sigma_R= 1$ kpc.   With the width of $\sigma_R=1$ kpc, 90\% of the seeded MSPs within the bulge are within 1.5 kpc from the center, which is approximately  within the estimated diffuse region of the bulge \citep{BldGer16}.  This second form was the MSP bulge density distribution used by \cite{Hooper13}.  Seeding the MSPs within the Galactic Bulge in a $7^\circ\times7^\circ$  ROI, we treated the MSPs in the simulation in the same manner as the MSPs from globular clusters.  We adjusted the \fermi point source threshold map for a viewing period of 45 months (GM13) and simulated a number of MSPs placed in the GC using the above density distributions.  We gave the bulge MSPs the same characteristics as the MSPs from the GD with the model parameters obtained for the TPC high-energy emission model.  We obtained the SEDs presented in Figure (\ref{fig:GCESpec}).  The solid red dots with red and black error bars represent the SED in Appendix A of GM13 associated with statistical and systematic uncertainties, respectively.  More recently, the \fermi team obtained an estimated SED of the GCE including a variety of systematic uncertainties in \cite{Ackermann12b}.  The SED of GM13 is contained within the uncertainties of this new SED as shown in Figure (15) of \cite{Ackermann12b}.  However, the new \fermi SED is harder than the SED used here in this study from GM13.   \cite{Hooper13} fitted the summed SEDs of 37 \fermi MSPs to obtain the best-fit values of $\Gamma=1.46$ and $E_{\rm cut}=3.3$ GeV.  Normalizing their SED to that of the GCE results with the following form
\eqs{\label{eq:hoop}
\left(E^2\,dN/dE\right)_{\rm H}\;=\;3.26\times 10^{-7}\,E^2\cdot E^{-1.46}\,e^{-E/3.30}\ \left({\rm GeV\cdot cm^{-2}\cdot s^{-1} } \right) }
\begin{figure}[h!]
\begin{center}
\includegraphics[trim=0.2in 0in 0in 0in, width=4.5in]{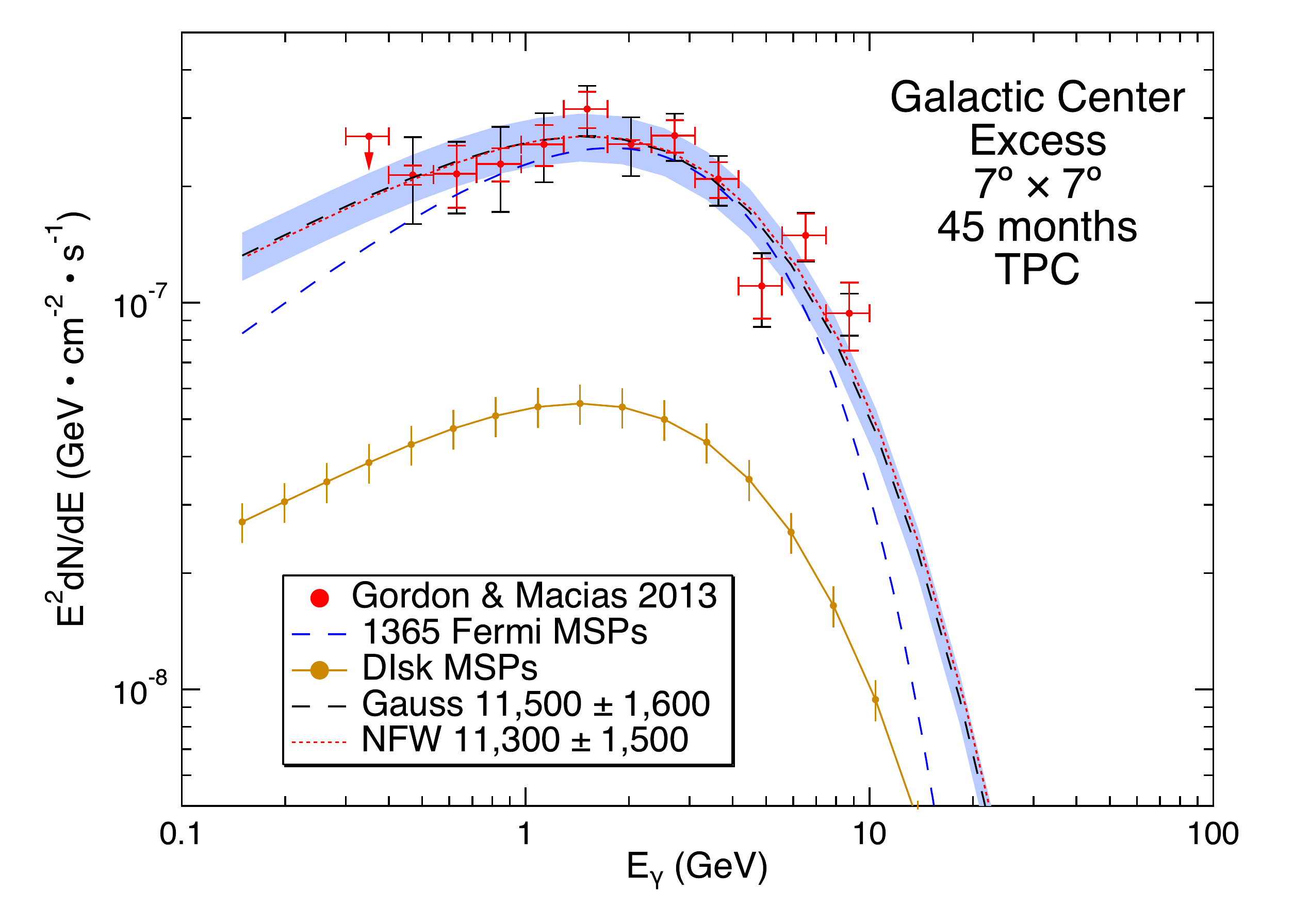}
\caption{The $\gamma$-ray power spectrum of the Galactic Center Excess.  The solid red dots with red (statistical) and black (systematic) error bars from Appendix A of \cite{Gordon13}.  The dashed blue curve is the average power spectrum of 39 MSPs in the 2PC after converting their fluxes from their estimated distances, placing them at the GC at 8.5 kpc, and multiplying the average by 1365 yielding a power spectrum  $E^2\,dN/dE\;=\;3.4\times 10^{-7}\,E^2\cdot E^{-1.29}\,e^{-E/2.5}\ \left({\rm GeV\cdot cm^{-2}\cdot s^{-1} } \right)$.  The solid brown curve with error bars represents the contribution of the MSPs from the Galactic Disk discussed within the text.   The dashed black and dotted light red curves essentially fall on top of each other represent the contributions of the the bulge MSPs from the two density distributions in Equations \ref{eq:NFW} and \ref{eq:Gau}, respectively, yielding similar power spectra given by the expression $E^2\,dN/dE\;=\;3.91\times 10^{-7}\,E^2\cdot E^{-1.49}\,e^{-E/3.22}\ \left({\rm GeV\cdot cm^{-2}\cdot s^{-1} } \right)$.  The light blue shaded region represents the 1 $\sigma$ confidence region of the simulations from either NFW or Gaussian density distributions. } 
\label{fig:GCESpec}
\end{center}
\end{figure}
We took 39 MSPs from the 2PC, converting their fluxes from their estimated distances from us and placed them at the GC at a distance of 8.5 kpc; then we averaged their fluxes and fitted the resulting SED.  In order to account for the SED of the GCE, we needed 1365 \fermi MSPs to account for the GCE shown by the dashed blue curve with the following form
\eqs{\label{eq:MSPfm}
\left(E^2\,dN/dE\right)_{\rm 2PC}\;=\;3.40\times 10^{-7}\,E^2\cdot E^{-1.29}\,e^{-E/2.50}\ \left({\rm GeV\cdot cm^{-2}\cdot s^{-1} } \right).}
  We indicate in the figure the predicted contribution from the previously discussed MSPs of the GD as shown with a solid brown curve with 1$\sigma$ error bars, which is about 10\% of the GCE.  Finally, the dashed black and dotted red curves that fall on top of each other correspond to the contribution of the  MSPs seeded in the bulge with the density distributions given by Equations \ref{eq:NFW} and \ref{eq:Gau}, respectively.  We found that about $11,500\pm 1,500$ bulge MSPs for either source distribution were required to account for the SED of the GCE with either density profile, resulting in the simulated SED with the fit of the form
 \eqs{\label{eq:bulge}
\left( E^2\,dN/dE\right)_{\rm sim}\;=\;2.9\times 10^{-7}\,E^2\cdot E^{-1.68}\,e^{-E/4.30}\ \left({\rm GeV\cdot cm^{-2}\cdot s^{-1} } \right).} 
These three SEDs, in the above Equations (\ref{eq:hoop}), (\ref{eq:MSPfm}), and (\ref{eq:bulge}), yield similar photon and energy fluxes of $\sim (1.4\pm0.3)\times 10^{-6}\ {\rm photons\cdot cm^{-2}\cdot s^{-1}}$ and $\sim (1.3\pm0.1)\times 10^{-9}\ {\rm erg\cdot cm^{-2}\cdot s^{-1}}$, respectively.

Since the assumed distributions also populate beyond the ROI, about 34,200 MSPs needed to be simulated to result in 11,500 MSPs within ROI for either NFW and Gaussian densities.  \cite{Hooper13} and GM13 indicated about 1,000 MSPs suffice, \cite{Wang05} suggested 6,000 MSPs are needed, while  \cite{Yuan14} found that 1 to 2 $\times 10^4$ MSPs with luminosities larger than $10^{32}$ erg/s are required to explain the GCE.  Using the prescription to estimate the number of MSPs in globular clusters given in Equation (\ref{eq:etaFm}), with values of $\langle\dot  E\rangle\;=\;(1.8\pm 0.7)\times 10^{34}$ erg/s and $\langle\eta\rangle\;=\; 0.08$ \citep{Abdo09Tuc} for the typical MSP, and assuming the distance to the GC is 8.5 kpc, Equations \ref{eq:hoop}, \ref{eq:MSPfm}, and \ref{eq:bulge} all yield very similar $\gamma$-ray luminosities of the GCE with the average luminosity $\langle L_\gamma\rangle \;=\; (1.13\pm 0.09) \times 10^{37}$ erg/s.  Using the above prescription in Equation \ref{eq:etaFm}, yields the number of bulge MSPs to be $8000\pm 3000$ in close agreement with the number predicted by our simulation.

In our simulation, the MSPs have smaller average luminosity than that of the detected \fermi MSPs.  Thus, our population synthesis suggests that along with the bright \fermi MSPs there is a population of MSPs that are much less luminous with smaller spin-down luminosities, requiring about 11,500 MSPs to explain the GCE.   We conclude that using the typical luminosities of the detected \fermi MSPs significantly underestimates the number of MSPs in the GC, suggesting that only about 1,000 MSPs cannot account for the GCE SED.  

\begin{figure}[h!]
\begin{center}
\includegraphics[trim=0.2in 0in 0in 0in, width=6.in]{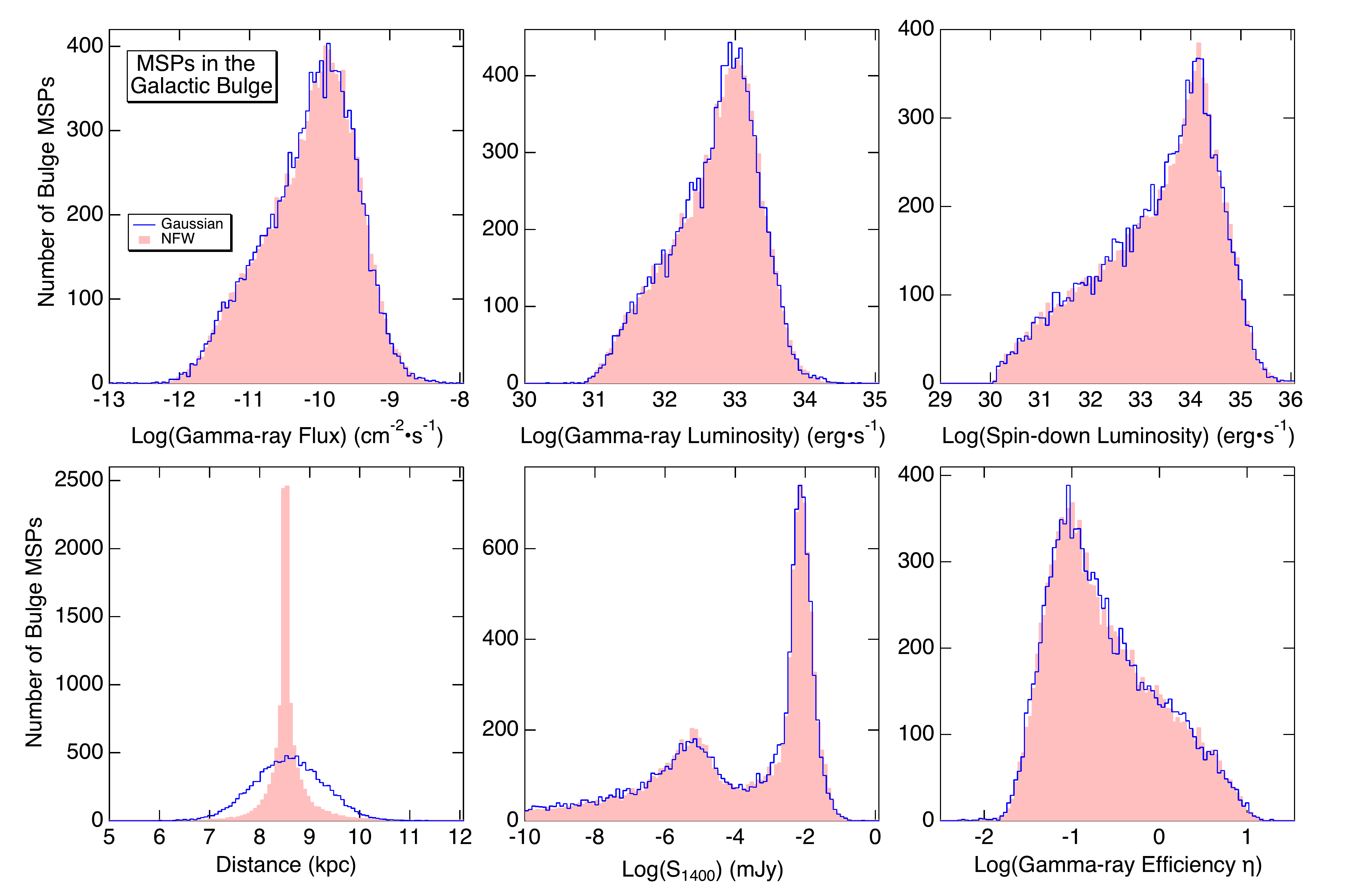}
\caption{Various distributions of indicated characteristics of the simulated bulge MSPs plotted for the Gaussian (blue) and NFW (red) density distributions of Eq.s \ref{eq:NFW} and \ref{eq:Gau}, respectively.}
\label{fig:Bulges}
\end{center}
\end{figure}

In Figure (\ref{fig:Bulges}), we plot various indicated characteristics of our simulated MSPs within the $7^\circ\times7^\circ$ ROI from the two bulge density profiles.  The luminosity distributions are broad distributions from $10^{31}$ to $10^{34}$ erg/s peaking around  $8\times 10^{32}$ erg/s with the spin-down luminosity distributions having a similar, but broader, shape peaking at $1.5 \times 10^{34}$ erg/s with $\gamma$-ray efficiencies  peaking at $0.06$ in fair agreement with \cite{Abdo09Tuc} (their average $\langle \eta\rangle=0.08$ is somewhat higher).  The distributions of the photon flux peaks at $5 \times10^{-10}\ {\rm photons\cdot cm^{-2}\cdot s^{-1}}$ dropping sharply above  $10^{-9}\ {\rm photons\cdot cm^{-2}\cdot s^{-1}}$.  All the simulated MSPs are below the \fermi point source threshold; note that the ROI is within the GC where the threshold is higher than average.  The furthest detected \fermi MSP is at a distance of about 3.5 kpc.  The distance distributions in Figure (\ref{fig:Bulges}) indicate that none of the simulated bulge MSPs is closer than 6 kpc.  As can be seen in the histograms of the logarithms of the radio luminosities $S_{1400}$ at 1400 MHz, there are some predicted to be above $10\ \mu$Jy that soon might be accessible with deeper radio searches.
For the double peak in the radio flux $S_{1400}$ histogram see the discussion of Figure (\ref{fig:47TucHis}).

\section{Discussion}
We have performed a population synthesis of radio and $\gamma$-ray MSPs from the Galactic Disk (GD) with the principal assumption that the MSP birthrate within the Galactic Disk is 4.5 MSPs per Myr as predicted in the work of SGH and agreeing with other studies (see SGH and the references therein).  The study of SGH was done before the launch of \fermi and so made predictions for one year of observing that \fermi would detect 12 radio-loud MSPs, those seen in the ten radio surveys in the study, and 33 to 40 radio-quiet, those not previously seen in radio surveys.  However, SGH indicated that these \fermi radio-quiet MSPs could be identified from their lack of variability, their spectral energy cutoffs, and that pulsations could be discovered through dedicated radio observations.  Indeed the discovery of \fermi MSPs has proceeded  in this manner with the exception, J1311-3430, detected in a \fermi blind search \cite{Ray13} and  J2339-0533 was detected through gamma-ray blind search after optical observations measured an orbital period \citep{Ray14}. 

Several improvements have been made over our previous study in SGH including the addition of three more radio surveys. We also tested three high-energy emission models, which involve an empirical $\gamma$-ray luminosity model with the exponents of $\dot P$ and $P$ and the normalization factor as free model parameters, rather than the theoretically motivated luminosity of the PSPC high-energy emission model as in SGH.  
We focused on the eleven-month \fermi first point source catalog 1FGL  \citep{Abdo1FGL}  to tune the six free model parameters associated with both the radio and $\gamma$-ray luminosity models.  We found eleven \fermi MSPs that were known as radio MSPs detected in the thirteen radio surveys used in this study and a total of 54 \fermi MSPs, a result that is in excellent agreement with the study of SGH.  
However, SGH implemented the PSPC high-energy model exclusively with the $\gamma$-ray emission arising from the entire open-field volume of the pulsar magnetosphere.  The \fermi MSP spectra being described by a power-law with an exponential cutoff clearly indicates high-altitude emission near or beyond the light cylinder.  The light curve shapes more strongly suggest narrow emission gaps in the outer magnetosphere.  The study of \cite{Johnson14} fitting the radio and $\gamma$-ray light curves,  found that only six of the MSPs in the 2PC preferred the PSPC high-energy model.  

In addition, we have developed a Markov Chain Monte Carlo (MCMC) code that efficiently explores the model parameter space by developing the likelihood with four and five 1-D histograms associated with radio and \fermi MSP characteristics, respectively.  The MCMC exploration of the parameter space of the luminosity models allow the estimation of the most likely values of the parameters and their 1 $\sigma$ uncertainties, resulting in values of the exponents of $P$ and $\dot P$ of $\alpha \sim -1.2 \pm 0.1$ and $\beta \sim 0.67 \pm 0.10$, respectively (see Table \ref{tab:paraunc}).    However, the common assumption that $L_\gamma$ is approximately equal to $\approx 8\%$ of the spin-down luminosity $\dot E$ requiring that $\alpha_\gamma\,=\,-3$ and $\beta_\gamma\,=\,1$, is ruled out by the MCMC simulation by well over 3 $\sigma$.  There is a very large scatter in the plot of the $\gamma$-ray luminosity versus $\dot E$ in Figure (\ref{fig:GamLum}) for the detected \fermi MSPs with differences of two orders of magnitude for MSPs with similar $\dot E$.  Here the $\dot E$ has been calculated for an orthogonal rotator ($\alpha\,=\,90^\circ$).  Such large variations are unlikely to arise from variations of the actual $f_\Omega$, which arise directly from the sky maps of the assumed high-energy emission models and are related to the average phase profiles from Equation (\ref{eq:AveFO}).  The uncertainties in $f_\Omega$ can account for a factor of a few as the simulations of all three high-energy emission models lead to a narrow distribution of $f_\Omega$.  It may very well be that the uncertainties associated with the $\gamma$-ray fluxes and distances of MSPs obtained from the electron density model NE2001 may indeed lead to larger variations of the inferred $\gamma$-ray luminosities.  On the other hand, it may be that the observed spread in $L_\gamma$ versus $\dot E$ of the detected \fermi MSPs is pointing to an alternative explanation, such as the possibility  that $L_\gamma$ has a stronger dependence on the magnetic inclination angle $\alpha$ that is seen in more realistic pulsar magnetospheres simulations\citep{Kalapotharakos16, Kalapotharakos18}.  While not a conclusion of this study, this result suggests that the next step in improving population syntheses requires the features of more realistic pulsar magnetospheres, such as the hybrid model with force-free electrodynamics within the light cylinder and dissipative regions (FIDO) beyond the light cylinder and Particle-in-Cell simulations \citep{Kalapotharakos14,Kalapotharakos18,Brambilla18}.

One of the goals in this study was to seek discriminating signatures among the TPC, OG, and PSPC high-energy models.  One of the drawbacks of the population synthesis is that ``detection" in the simulation is sensitive to the average $\gamma$-ray flux and not sensitive to light curve features of the MSP.  Therefore, we included the PSPC along with the TPC and OG high-energy emission model in this study to explore signatures that might distinguish these models.  Yet, no clear distinctions were found among the high-energy emission models in the comparison with the 1-D histograms in Figure (\ref{fig:Fermi}) of detected MSP features.  However, there are large regions in the viewing angle $\zeta$ versus inclination angle $\alpha$ in Figure (\ref{fig:Foms}) and in Figure (\ref{fig:ZetaAlpha}) with $\alpha\, <\, 50^\circ$ and $\zeta\,<\, 60^\circ$ that exclude possible viewing geometries for the OG high-energy emission model.  Only two MSPs in the study of \cite{Johnson14} whose resulting $\zeta$ and $\alpha$ lie in this excluded region, one for the TPC and one for the PSPC, fall in this region making it difficult to find a distinguishing feature of these high-energy emission models.  The regions of smaller $f_\Omega$ are favored by the simulations as the smaller the $f_\Omega$, the larger the $\gamma$-ray flux.  Of the five MSPs that were fit within the PSPC model, only one falls in this region of small $f_\Omega$.
However, this is also true for the TPC and OG models where the resulting $\zeta$ and $\alpha$ from the fits with $\zeta\,>\, 60^\circ$ and $\alpha\, <\,30^\circ$ do not tend to fall in regions of small $f_\Omega$, and therefore these fits lead to MSPs with very small radio fluxes.  We conclude that future fitting of radio and $\gamma$-ray light curves not only should consider the shape of the light curves but also need to  take into account  the corresponding fluxes, and therefore the absolute profile intensities.

The population synthesis presented here suggests that there are, aside from the \fermi point sources that have been discovered to be $\gamma$-ray MSPs, a comparable number of \fermi point sources that are $\gamma$-ray MSPs with much weaker radio fluxes $S_{1400}$.  To tune our free parameters of the $\gamma$-ray luminosity model, we imposed constraints on the simulated \fermi MSPs in order that they have radio fluxes $S_{1400}$ above $30\ \mu$Jy and dispersion measure DM greater than 2.5 $\rm pc/cm^3$, corresponding to the lower limits for radio detection and subsequent pulsation detections of our group of 54  \fermi MSPs in the 1FGL.  Figure (\ref{fig:S1400DM}) suggests that there are about an equal number of MSP candidates detected as \fermi point sources within the 1FGL  catalog with significant $\gamma$-ray fluxes.  However, these MSPs have radio fluxes $S_{1400} \,<\, 30\ \mu$Jy with about 30 MSPs with $0.1\ \mu{\rm Jy}\, <\, S_{1400}\, <\, 30\ \mu{\rm Jy}$.  Our simulations conclude that many more MSPs can be discovered through deeper pointed radio observations with larger radio telescopes.

Having established the six free parameters associated with the radio and $\gamma$-ray luminosity models, we explored how well the population synthesis predictions agree with the number of MSPs discovered in other point source catalogs listed in Table \ref{tab:Proj}.  For each of these point source catalogs, we adjusted only the factor multiplying the \fermi point source threshold map, using the accepted MCMC steps after a burn-in period to obtain 1 $\sigma$ uncertainties.  As indicated in Section \ref{sec:selgrp}, we selected 54 \fermi MSPs detected as point sources with radio pulsations detected in dedicated deep radio pointed observations.  More recent observations with FAST \citep{Nan2006}, Einstein@Home \citep{Wu2018} and observations at various other radio wavelengths have discovered more MSPs in \fermi point source catalogs.  Our simulation predicts these as part of the radio-weak $\gamma$-ray MSPs with the numbers listed in Table \ref{tab:Proj}.   In the cases of 2FGL \citep{Nolan12} and 3FGL \citep{Acero15}, the number of MSPs predicted by the model of about 80 and 110 is about 30\% larger than those discovered so far with recently updated numbers of 54, 64, and 87, for the 1FGL, 2FGL, and 3FGL catalogs.  Our predictions for total number of $\gamma$-ray MSPs are about 100, 150, and 215 for 1FGL, 2FGL, and 3FGL.  These predictions compare favorably to recent detected total numbers of MSPs of 64, 74, and 98, for the corresponding catalogs.  However, in these simulations, we did not pay careful attention to the $S_{1400}$ and DM limits of the pointed radio observations of sources in these catalogs, which may account for these differences.  It is also clear especially in the cases of 2FGL and 3FGL, the catalogs of \fermi MSPs detected as point sources are incomplete.  It is also likely that some of the MSPs in these later catalogs remain to be discovered.  What clearly stands in contrast, is the bright $\gamma$-ray source list catalog BSL \citep{Abdo09} for 3 months of observing, where the code predicts about 100\% more MSPs than the 13 discovered.  This can be explained by the BSL threshold test statistic TS greater than 100 while the subsequent catalogs implemented a threshold TS above 25.  The simulation predicts that about 120 and 170 MSPs will be detected by \fermi in 5 and 10 years, respectively, advocating for \fermi to further increase the population of $\gamma$-ray MSPs.

With our population model tuned to reproduce the general characteristics of measured properties of MSPs from the GD, we find that the MSPs from the GD contribute less than 4\% and 2\% to the diffuse emission of the $\gamma$-ray diffuse background in the Galactic Plane and out of the plane, respectively.  The study of \cite{Faucher10} conclude a higher contribution of 5 to 15\% to the high-latitude $\gamma$-ray background.   On the other hand, \cite{Calore14} find that above 2 GeV, MSPs contribute at most 0.9\% of the isotropic diffuse $\gamma$-ray background.  However, their study was based on fitting the distributions of observed MSPs to make conclusions regarding the distributions of the undetected MSPs contributing to the diffuse $\gamma$-ray emission. Our conclusion depends crucially on our central assumption that the birthrate of MSPs is 4.5 per Myr in agreement with SGH and other studies (see SGH for references therein).  

Given the increasing interest in identifying the source of the $\gamma$-ray Galactic Center Excess (GCE), we also simulated a population of MSPs within the Galactic Bulge with the same properties determined for the MSPs (both detected and undetected) from the GD.  With the spectral shape and intensity of the excess estimated, two main sources have been discussed in the literature, the annihilation associated with a dark matter particle or a yet undetected population of MSPs within the bulge. We find that the simulation produces a population of MSPs from the GD that contributes only about 10\% to GC and so cannot account for the GCE (see Figure \ref{fig:GCESpec}).  We agree that the group of MSPs that \fermi has detected has an average spectral shape that agrees very well with the GCE as presented by \cite{Gordon13}.  We also agree that if somewhat over a thousand of these \fermi MSPs would be placed at the GC that they would account for the spectral intensity of the GCE.  However, the simulation of the MSPs from the GD requires a power law in the magnetic field producing MSPs with a wide range in $\dot E$ and therefore a luminosity function that extends to much lower values than those detected.  So along with the high-$\dot E$ \fermi MSPs, our simulation includes all those with lower $\dot E$ that are below the \fermi detection threshold as well as below radio detection limits.  Therefore, we conclude that about 11,500 MSPs are required in the inner GC within the $7^\circ\times7^\circ$ region of interest.  

The assumed origin of these MSPs in the Galactic bulge is the inspiraling of globular clusters in a dynamical friction model as treated by various groups and revisited recently by \cite{Gnedin14} who estimated $(2 - 6) \times 10^7\ M_\odot$ in globular clusters were dragged into a region of 3 - 5 pc from the GC.  \cite{Gnedin97} list the masses of 119 globular clusters with an average mass of $3.8 \times 10^5\ M_\odot$ per cluster, suggesting that 50 to 160 globular clusters were captured by the GC.  If our estimates are correct, as we concluded in Table \ref{tab:Globular}, then on average there are 100 MSPs per cluster leading to a number between 5,000 and 16,000 MSPs in the GC, agreeing very well with the prediction of the population simulation of the GC.

The average luminosity of the simulated MSPs in the GC was estimated to be $\langle L_\gamma\rangle \;=\; (1.08\pm0.14) \times 10^{33}$ erg/s, which was in fairly good agreement with the estimated average luminosity of the MSPs in globular clusters in Table \ref{tab:Globular} as measured by the \fermi team in \cite{Abdo10}.  This means luminosity is a factor of 10 smaller than the one recently obtained by \cite{Hooper16} from a reanalysis of \fermi detections of globular clusters correlating stellar encounter rates and visible luminosity with the number of MSPs.  Their analysis seems to underestimate the number of MSPs in the clusters by making them significantly more luminous than the luminosity derived from our simulation that agrees with the \fermi team's assessment of the globular clusters as in \cite{Abdo10}.   \cite{Hooper16} would conclude that most of the $\gamma$-ray luminosity of 47 Tuc is the result of a single luminous MSP.   It might be possible that the luminosities of MSPs in globular clusters are different from those in the GD, as we have assumed.  Yet, 25 radio MSPs are known \citep{Pan16} to exist in the cluster.  Our population synthesis of MSPs from the Galactic Disk reproduces fairly well the radio and $\gamma$-ray characteristics of these pulsars, predicting reasonable numbers of \fermi MSPs for various observing periods.  Applying the model to globular clusters in \cite{Abdo10} reproduces the number of MSPs estimated by the \fermi team.  

From Figure (\ref{fig:Bulges}), we find that the simulation suggests that a large group of the MSPs in the GC has radio fluxes $S_{1400}$ between 1 and $10\ \mu$Jy.  These fluxes are well below many deep searches of \fermi unassociated sources.  The search with the Effelsberg telescope \citep{Barr13} did not get below $100\ \mu$Jy, while some of the searches from the \fermi Pulsar Search Consortium \citep{Ray12} have reached sensitivities below $10\ \mu$Jy.  We anticipate that deeper searches and larger radio telescopes like the new 70-meter Deep Space Network antenna in Canberra \citep{Lemley16} with its high frequency receiver (18-28 GHz) will soon detect MSPs within the Galactic Bulge.  

\acknowledgments{We greatly appreciate the lengthy conversations with Cole Miller regarding the development of the MCMC code.  We thank the referee for carefully reading the manuscript and providing many helpful comments that led to its improvement.  This research was
supported in part by the Howard Hughes Medical Institute through the Undergraduate Science Education Program.  We also are grateful for the generous support of the Michigan Space Grant Consortium, the National Science Foundation under grant Nos. REU: PHY/DMR-1004811, and RUI: AST-1009731, the \fermi Guest Investigator Program Cycle 3 under grant No. NNH09ZDA001N-FERMI3, and the NASA Astrophysics Theory and Fundamental Physics Program under the grant No. NNX13AO12G.  }

\bibliographystyle{aasjournal}

\end{document}